\documentclass[preprint,aps]{revtex4}
\usepackage{amsmath}
\usepackage{graphicx}
\usepackage{epsfig}
\newlength{\upit}\upit=0.1truein
\newcommand{\raiser}[1]{\raisebox{\upit}[0cm][0cm]{#1}}
\newcommand{\ltappr}{{{\lower4pt\hbox{$<$} } \atop \widetilde{ \ \ \ }}}
\newlength{\bxwidth}\bxwidth=1.5 truein
\newcommand\frm[1]{\epsfig{file=#1,width=\bxwidth}}
\begin{document}
\newcommand{\dg}{^{\dagger }}
\newcommand{\si}{\sigma}
\newcommand{\rarrow}{\rightarrow}
\def\fig#1#2{\includegraphics[height=#1]{#2}}
\def\figx#1#2{\includegraphics[width=#1]{#2}}
\newlength{\figwidth}
\figwidth=10cm
\newlength{\shift}
\shift=-0.2cm
\newcommand{\fg}[3]
{
\begin{figure}[ht]
\vspace*{-0cm}
\[
\includegraphics[width=\figwidth]{#1}
\]
\vskip -0.2cm
\caption{\label{#2}
\small#3
}
\end{figure}}
\newcommand{\fgb}[3]
{
\begin{figure}[b]
\vskip 0.0cm
\[
\includegraphics[width=\figwidth]{#1}
\]
\vskip -0.2cm
\caption{\label{#2}
\small#3
}
\end{figure}}

\newcommand \bea {\begin{eqnarray} }
\newcommand \eea {\end{eqnarray}}
\newcommand{\bk}{{\bf{k}}}
\newcommand{\bx}{{\bf{x}}}

\title{Quantum replica approach to the under-screened Kondo model}

\author{P. Coleman$^{1}$ \& I. Paul$^{2}$}

\affiliation{$^{1}$ Center for Materials Theory,
Rutgers University, Piscataway, NJ 08855, U.S.A. }  

\affiliation{$^{2}$ SPhT, L'Orme des Merisiers, CEA-Saclay, 91191
Gif-sur-Yvette France.}

\begin{abstract}
We extend the Schwinger boson large $N$ treatment of the under-screened
Kondo model in a way that correctly captures the finite elastic phase
shift in the singular Fermi liquid.  The new feature of the approach,
is the introduction of a flavor quantum number with $K$ possible
values, associated with the Schwinger boson representation. The large
$N$ limit is taken maintaining the ratio $k=K/N$ fixed.  
This approach differs from previous approaches, in that we do not
explicitly enforce a constraint on the spin representation of the 
Schwinger bosons.  Instead, the 
energetics of the Kondo model cause the bosonic degrees of freedom to ``self assemble'' into
a ground-state in which the spins of $K$ bosons and 
$N-K$ conduction electrons  are antisymmetrically arranged into a
Kondo singlet.  With this device, the large $N$ limit can be taken,
in such a way that a 
fraction $K/N$ of the Abrikosov Suhl resonance is immersed inside the Fermi sea.
We show how this method can be used to model the full energy
dependence of the singular Abrikosov Suhl resonance in the
under-screened Kondo model and the field-dependent
magnetization. 
\end{abstract}
%
\maketitle
%
\section{Introduction}\label{}

The work in this paper is motivated by the physics of heavy electron
materials.  These materials are the focus of renewed attention, in
part because of the opportunity they present to understand the physics
of matter near a quantum critical
point\cite{questions,stewartrmp,varma,qcp,cox}.  One of the
unexplained properties of heavy electron quantum criticality, is that
the characteristic temperature scale of heavy electron Fermi liquid is
driven to zero at the quantum critical
point\cite{hvl,grosche,devisser,knebel,gegenwart}.  When either the
paramagnet or antiferromagnetic heavy electron phase is warmed above
this temperature scale, it enters a ``non-Fermi liquid'' phase.  The
standard ``Moriya-Hertz'' theory\cite{moriya,hertz} of quantum
magnetism is unable to explain the divergence of the heavy electron
mass in these three dimensional materials.  Many other aspects, such
as the appearance of $E/T$ and $H/T$ scaling in physical
properties\cite{schroeder00,gegenwart03}, the development of a
quasi-linear resistivity and the tentative observation of a jump in
the Hall constant at the quantum critical point\cite{silke04} suggest
that we have not yet found the correct mean-field theory for the
development of magnetism in these systems.

These considerations motivate  a renewed effort to find 
the correct mean-field theory that 
spans the quantum critical point between the antiferromagnetically ordered Kondo
lattice and fully screened Kondo lattice paramagnet.  Existing mean-field treatments
of the heavy electron paramagnet describe the spin degree of freedom as a
fermionic bilinear\cite{abrikosov,read,auerbach}, and while these methods provide an adequate
description of the formation of the heavy electron bands at low
temperatures,  they are ill-suited for a description of the antiferromagnetically ordered
state. 

This suggests that further progress  may require 
a bosonic mean-field description of both the 
Kondo impurity and lattice model.  Bosonic spin representations
have the advantage that they are naturally suited to the description of
antiferromagnetism in the Kondo lattice\cite{arovas}. In these
approaches, the spin rotation group is generalized from $SU(2)$ to $SU (N)$,
providing $1/N$ as a small expansion parameter.  
The hard part of the
problem is to capture the screening physics of the Kondo effect using
the bosonic spin description. 
A first step in this direction was made by 
Parcollet and Georges\cite{parcollet97a},  who argued that
in order to produce a Kondo singlet in an SU($N$) approach, one needs
to introduce a multi-channel Kondo model, in which the number of
screening channels $F$ grows with $N$. In their approach, it became
possible to describe the fully screened Kondo singlet by choosing the
number of screening channels equal to the number of bosons in the spin
representation, $F=2S$.  One of the difficulties  encountered
in this work, is that the localized moment occupies only
$1/N$ th of the singlet, giving rise to a vanishingly small elastic scattering phase
shift of $\pi/N$\cite{parcollet97a}.

In a more recent return to the Schwinger boson description of the
Kondo model\cite{colemanpepin03}, it was shown that  a controlled
large $N$ treatment of the under-screened Kondo model (UKM) can
actually be obtained using a single-channel Kondo model.  This method
captures the partial screening from spin $S$ to spin $S-1/2$,
revealing that UKM 
is a singular Fermi
liquid, where the slow logarithmic decoupling of the partially
screened moment\cite{mattis,noz} generates a 
a singular logarithmic dependence of the scattering phase
shift\cite{borda}.  
This is manifested by a singular
divergence
in the specific heat coefficient $C_{V}/T\sim 1/ (T \ln ^{4}
(T_{K}/T)) $
and differential magnetic susceptibility $\chi (B)\sim 1/
(B \ln ^{2} (T_{K}/B))$\cite{exact1,exact2}.

One of the interesting aspects of the under-screened Kondo
model, is that it  displays a field tuned Fermi temperature
which rises linearly with field. This is a feature found to be present 
at a heavy electron quantum critical point\cite{gegenwart03}.
However, this approach still leads to a $\pi/N$ phase shift
which vanishes in the large $N$ limit.

In this paper we continue this earlier work, showing how the phase
shift problem is solved by introducing $K$ replicas of the Schwinger
boson spin, writing
\[
\vec{S} = \sum _{\mu =1}^{kN}\vec{ S} (\mu)
\]
where the number of replicas $K=kN$ is scaled with N. 
This technique preserves a finite fraction  $k$ of the
impurity spin inside the Kondo singlet, and the 
Abrikosov Suhl resonance which develops is now immersed beneath the
Fermi sea with an  elastic phase shift  $\pi K/N$. 
Our results clearly show the formation of a singular
Abrikosov Suhl resonance with a finite phase shift, but 
at present they do not extend to the fully
screened Kondo model. We shall discuss at the end of this article 
how a future fusion of our
method with the multi-channel approach may indeed provide a viable
Fermi liquid description with a finite phase shift. 

\section{The Model}\label{}

Our starting point is the Kondo model, which is written
\begin{eqnarray}\label{start}
H&=& \sum_{k\alpha } \epsilon _{k}c\dg _{k\sigma }c_{k\sigma
}+ J \vec S\cdot \psi \dg _{\alpha} \vec \sigma_{\alpha \beta} 
\psi _{\beta}
.
\end{eqnarray}
where $S$ denotes a spin $S>\frac{1}{2}$, $c\dg _{k \alpha }$ creates
a conduction electron with wave vector $k$, spin component $\alpha$,
$\psi\dg _{\alpha }= \sum_{k}c\dg_{k\alpha }$ creates a conduction
electron at the impurity site.   
Our next step is to reformulate the UKM
as an SU(N) invariant Coqblin Schrieffer
model, which enables us to carry out a large $N$ expansion of the
physics. We write
\begin{eqnarray}\label{H1}
H&=& \sum_{k\alpha } \epsilon _{k}c\dg _{k\alpha }c_{k\alpha
}+\frac{J}{N}\sum_{\alpha \beta }\hat S_{\alpha \beta }\psi \dg _{\beta }
\psi _{\alpha }
\end{eqnarray}
where the spin indices run over $N$ independent values $\alpha ,\beta
\in (1,N)$.  The new feature in our treatment, is the introduction of
spin replicas. The  spin operator is written as a sum of $K=kN$
replicas
as follows:
\[
\hat S_{\alpha \beta}= \sum _{\mu=1}^{kN}b\dg _{\alpha \mu}b_{\beta
\mu}
\]
We shall consider the case where there are $2S$ bosons of each flavor, 
\[
n_{b\mu} = \sum_{\alpha =1}^{N}b\dg _{\alpha \mu}b _{\alpha \mu} = 2S\equiv
N\tilde{S}.
\]
where $\tilde{S}= 2S/N$ is kept finite as $N\rightarrow \infty$. 
The hope behind this 
approach, is that by retaining the bosonic character
of the spin, we should later be able to adapt this method to describe 
magnetic behavior in a Kondo lattice. 
A multi-channel, rather than a multi-flavor
formulation of the above model, has previously been treated within an
integral equation formalism\cite{parcollet97a} and a single-flavor
version was considered in \cite{colemanpepin03}.

To understand the reasoning behind the introduction of a flavor index,
it is  helpful to consider the strong-coupling limit of this model,
where the dispersion of the conduction electrons is ignored. 
When the number of flavors $K=1$, only one boson can bind with the
conduction electrons to form a singlet, and the remaining $2S-1$ bosons
form a decoupled local moment, as shown in Fig. \ref{fig1} (a).  When
$K$ exceeds unity, it becomes possible for $K$ bosons, each of different
flavor, to antisymmetrize and form a singlet with $N-K$ conduction
electrons, leaving behind $K$ decoupled spins, each with
$S^{*}=S-1/2$, as shown in \ref{fig1} (b).

\figwidth=8cm
\fg{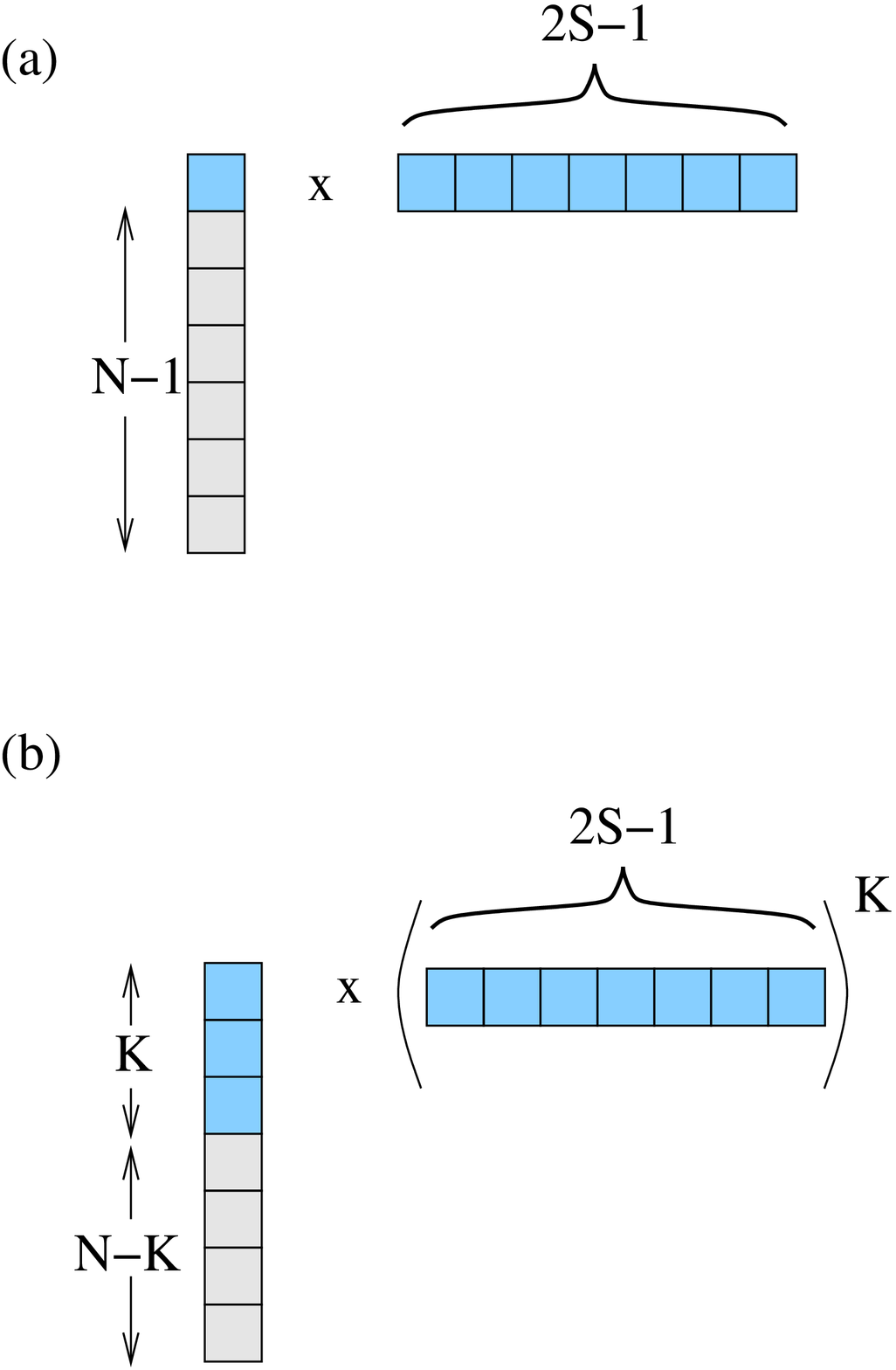}{fig1}
{Young tableaux which illustrate the strong coupling ground state of (a) the single
flavor model, with $K=1$ and (b) the multi-flavor model with $K>1$.
In (a), only one boson binds to the conduction electrons to form a
singlet. In (b),  $K$ bosons of different flavor can antisymmetrize
with each other, to form a singlet with $N-K$ conduction
electrons. Weak coupling models flow to this strong coupling limit,
giving rise to an elastic phase shift $\delta = \pi K/N$.}

The corresponding singlet ground-state is given by 
\begin{equation}\label{}
\vert \Psi \rangle = \int d\eta _{1}\dots d\eta _{K}\prod _{\sigma =1}^{N}
\left(b\dg _{\sigma \mu}\eta _{\mu}+ c\dg _{\sigma } \right)\vert S^{*}_{K} \rangle 
\end{equation}
where we have introduced $K$ Grassman numbers so that the composite
operator $f\dg _{\sigma }= \sum_{\mu=1}^{K}b\dg _{\sigma \mu}\eta
_{\mu}$ has the exchange symmetry of a fermion.  
The ket  $\vert S^{*}_{K}\rangle $ denotes any  state formed from $2S^{*}$
Schwinger bosons of each flavor (e.g. $\vert S^{*}_{K}\rangle =
\prod_{a=1}^{K} (b\dg _{\uparrow a})^{2S^{*}}\vert 0\rangle $ ).
Carrying out the
Grassman integral, this becomes
\[
\vert \Psi \rangle = \epsilon_{\alpha _{1}\dots \alpha _{N}}\epsilon_{^{_{\mu_{1}\dots \mu_{K}}}}
b\dg _{\alpha _{1}\mu_{1}}
\dots
b\dg _{\alpha _{K}\mu_{K}}c\dg _{\alpha _{K+1}}\dots c\dg _{\alpha
_{N}}\vert S^{*}_{K} \rangle .
\]
We shall consider a large $N$ limit in which the ratio $k=K/N$ of flavor
to spin degeneracy remains fixed, 
so that for instance, by considering $k=1/3$, 
we make it possible for
the Schwinger bosons to form a one-third filled singlet ground-state,
even in the large $N$ limit.   
The condition that $k$ is fixed as $N\rightarrow \infty $ guarantees
that the impurity Free energy grows as $O (N)$, which is the condition 
for a controlled large $N$ expansion. 

\section{Integral equations for the large $N$ limit.}\label{}

Our first step is to 
cast the partition function as a path integral and  factorize the interaction
\begin{equation}\label{}
H_{I}\rightarrow 
\sum_{\mu=1}^{K}
\left[
\frac{1}{\sqrt{N}}
\bar \phi_{\mu}\  
\left( b\dg _{\sigma \mu }\psi _{\sigma } \right)
+\frac{1}{\sqrt{N}}
\left(\psi \dg_{\sigma }b_{\sigma \mu} \right)
\phi _{\mu}
-\frac{1}{J}
\bar \phi_{\mu} \phi_{\mu}
\right]
.
\end{equation}
The Lagrangian for this model is then written
\begin{equation}\label{}
S = \int _{0}^{\beta }d\tau  \sum_{k,\sigma }c\dg
_{k\si}\left(\partial_{\tau }+\epsilon_{k} \right) c_{k\si} + \sum_{\sigma
\mu}b\dg _{\sigma \mu}\left(\partial_{\tau }+\lambda \right)b_{\sigma
\mu}
- 2 S K
\lambda + \int _{0}^{\beta }d\tau  H_{I}
\end{equation}
where the $\lambda$ field imposes the constraint ${\cal  N}_{b}= \sum_{\sigma
\mu }b\dg _{\sigma \mu}b _{\sigma \mu}= 2 S K$.  The method we now follow is
closely analogous to that of Parcollet and Georges\cite{parcollet97a}.
First, we integrate out the
bosons, writing  $S=S_{b} (\lambda) + S_{K}$, where
\[
S_{b} (\lambda) = NK  \log[1- e^{-\beta \lambda}] - 2S K \lambda/T.
\]
is the part of the action describing the free boson and 
\begin{eqnarray}\label{}
S_{K} &=& \int _{0}^{\beta }d\tau  
\left[ \sum_{k,\sigma }c\dg
_{k\si}\left(\partial_{\tau }+\epsilon_{k} \right)c_{k\si} -\frac{1}{J}
\bar \phi_{\mu} \phi_{\mu} 
\right]+S_{I}\cr
S_{I}&=& -\frac{1}{\beta N}\sum_{\sigma, \mu}\int_{0}^{\beta } d\tau d\tau '
 \psi \dg _{\sigma } (\tau ) \psi _{\sigma } (\tau' )
D_{o} (\tau -\tau ')
\bar \phi _{\mu} (\tau ')\phi_{\mu} (\tau )
\end{eqnarray}
is the ``Kondo'' contribution to the action, 
where $D_{0} (i\nu_{n})= \frac{1}{i \nu_{n}-\lambda}$.

If we carry out a Hubbard Stratonovich decoupling of $S_{I}$, to
obtain
\begin{eqnarray}\label{}
S_{I}  &=& 
\frac{1}{\beta}\int_{0}^{\beta } d\tau d\tau ' \left[\sum_{\sigma }
\psi \dg(\tau )
\Sigma (\tau -\tau ')\psi (\tau' ) + \sum_{\mu}\bar  \phi _{\mu}
(\tau )
\Pi (\tau -\tau ') 
\phi _{\mu} (\tau ') \right.
\\ &+& \left. 
N \Pi (\tau -\tau ') \frac{1}{D_{0} (\tau -\tau ')} \Sigma (\tau '-\tau )
 \right]
\end{eqnarray}
As $N$ becomes large, each term in $S_{K}$ grows extensively as $\sim N$, so
that in the large $N$ limit, the saddle point of this action is
expected to saturate the path-integral, giving an essentially exact
solution to the problem in the large $N$ limit. 
The Hubbard Stratonovich transformation has in essence, replaced
\begin{eqnarray}\label{replace}
-
\frac{1}{N}
 \sum_{\mu}D_0 (\tau -\tau ')\langle  \bar \phi _{\mu} (\tau')\phi _{\mu} (\tau )
\rangle 
&\rightarrow& \Sigma (\tau -\tau ')\cr
-
\frac{1}{N}\sum_{\si} D_0 (\tau' -\tau) 
\langle \psi\dg _{\si} (\tau' ) \bar \psi_{\sigma} (\tau )\rangle 
&\rightarrow & \Pi (\tau -\tau ')
\end{eqnarray}
These replacements become identities in the large $N$ limit. 
To see this directly, we integrate out the Fermions to obtain the
effective
Free energy:
\begin{eqnarray}\label{freeenergy}
F_{eff} &=& T S_{b} (\lambda) - K T {\rm  Tr}\ln \left[-\frac{1}{J}+ \Pi \right] 
- N T {\rm  Tr}\ln \left[- g_{o}^{-1}+\Sigma \right] 
 \cr &+& \frac{N}{\beta^2 }\int d\tau d\tau '
\Pi (\tau '-\tau )\frac{1}{D_{0} (\tau -\tau ')}\Sigma (\tau -\tau ').
\end{eqnarray}
This is the starting point for producing our mean-field equations.  If
we differentiate w.r.t. $\Pi (\tau '-\tau )$ we obtain
\begin{equation}\label{}
\frac{1}{\beta}\overbrace {\left(\frac{K}{\frac{1}{J}- \Pi 
} \right)(\tau -\tau ')}^{K{\cal J} (\tau -\tau ')} + \frac{N}{D_{0} (\tau -\tau ')}
\Sigma (\tau -\tau ')=0
\end{equation}
or 
\[
\Sigma (\tau -\tau ')= - \frac{K}{N \beta}D_{0} (\tau -\tau ') 
{\cal J} (\tau -\tau ')
\]
where 
\[
{\cal J} (\omega) =  \frac{1}{\frac{1}{J_{0}}- \Pi (\omega)}
\]
Similarly, differentiating w.r.t. $\Sigma (\tau '-\tau )$ we obtain
\begin{equation}\label{}
\frac{1}{\beta} \overbrace {\left(\frac{1}{g_{o}^{-1}- \Sigma} \right) (\tau -\tau'
)}^{g (\tau -\tau ')} +  \Pi (\tau
,\tau ') \frac{1}{D_{0} (\tau -\tau ')}=0,
\end{equation}
or 
\begin{equation}\label{}
\Pi (\tau -\tau ') = - \frac{1}{\beta}D_{0} (\tau '-\tau )
 g (\tau -\tau '),
\end{equation}
where 
\begin{equation}\label{}
g (\omega) = \frac{1}{g_{0}^{-1} (\omega)- \Sigma (\omega)}.
\end{equation}

If we convert the integral equations to Matsubara summations, we
obtain
\begin{eqnarray}\label{}
\Sigma (i \omega_{n}) &=& -k T
\sum _{r}\frac{1}{i  \nu_{r} -
\lambda} {\cal J} ( i \omega_{n}- i \nu_{r})\cr
\Pi (i \omega_{n}) &=& -T \sum _{r}\frac{1}{i  \nu_{r} -
\lambda} g ( i \omega_{n}+ i \nu_{r}).
\end{eqnarray}
These equations  governing the large $N$ limit can simply understood 
diagrammatically as ``NCA'' or non-crossing diagrams, as shown in
Fig. \ref{daig}.  
\figwidth=12cm

\fg{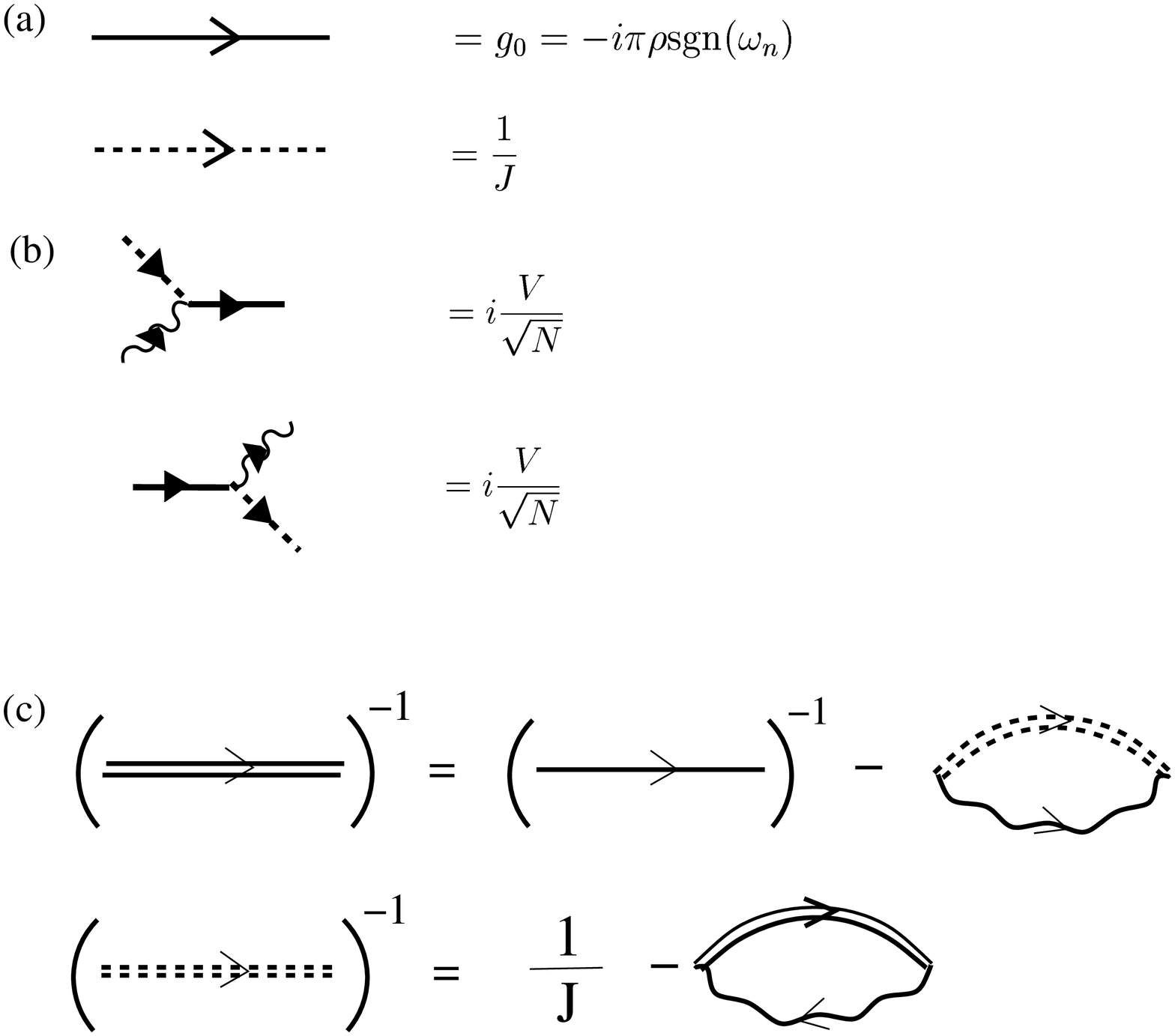}{daig}{Diagramatic representation of large $N$
equations. (a) Propagators for the conduction and $\phi $ fermion; (b)
vertex between particles (c) self-consistent ``NCA''diagrams for the
self-energies. 
}
\figwidth=8cm

Carrying out the Matsubara summations, we obtain
\begin{eqnarray}\label{core1}
\Sigma (z)&=& kn_{b}\ {\cal  J} (z-\lambda)+ k\int \frac{dy}{\pi}\frac{1- f (y)}{z-y-\lambda}Im
{\cal J}
(y-i\delta )\\\label{core2}
\Pi (z)&=& n_{b}\  g (z+\lambda)+\int \frac{dy}{\pi}\frac{ f (y)}{z-y+\lambda}Im g
(y-i\delta )
\end{eqnarray}
where $n_{b}= 1/ (e^{\beta \lambda}-1)= n (\lambda)$ and $K/N = k$.  
These integral equations can be solved simply by numerical iteration.

\section{Frequency dependent t-matrix }

The above integral equations (\ref{core1},\ref{core2}) were solved  by an iterative numerical
procedure. One of the key quantities of interest, is the conduction
electron phase shift, given by
\[
\delta _{c}= {\rm  Im }\ln \left[1 - i \pi \rho 
\Sigma (0 - i \delta ) \right],
\]
Under the assumption that $K=kN$ bosons are bound into the singlet,
Friedel's sum rule determines that the conduction electron phase shift
will be $\delta_{c}= k \pi$.  Our numerical results
(Fig. \ref{figphase}) confirm this result.
\figwidth=12cm
\fg{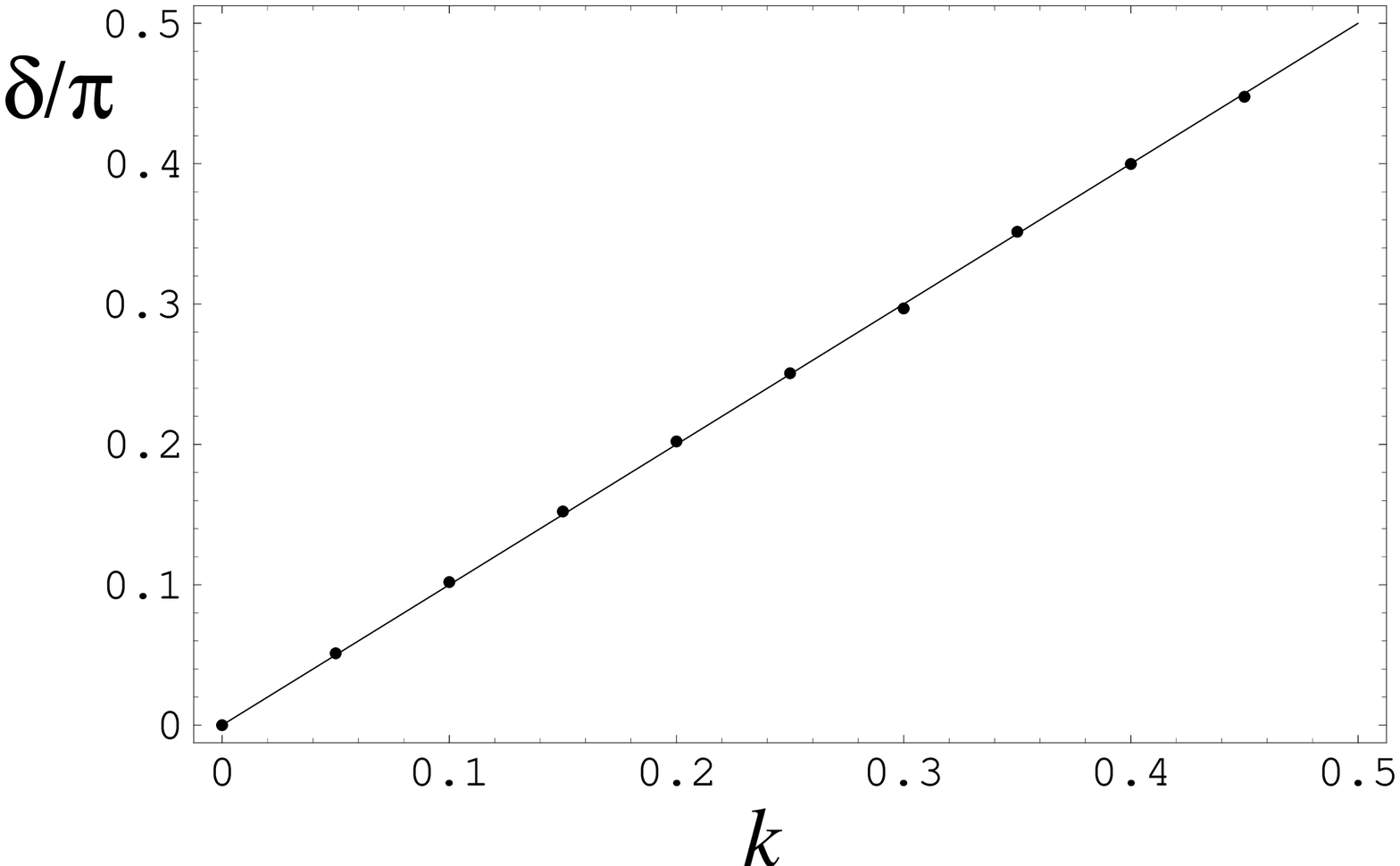}{figphase}{Showing the numerically computed
dependence of the conduction electron phase shift on $k$. These phase shift
were actually computed from at zero temperature 
from the zero-field limit of the finite temperature
conduction electron Green's functions. (See section VI)
}

To get an approximate understanding of the numerical results, it
sufficient to carry out the first iteration. If we assume $g^{(0)} (\omega-i
\delta )= i \pi \rho $, where $\rho $ is the conduction density of
states, then the first order approximation for $\Pi (\nu)$ is
\begin{eqnarray}\label{}
\Pi^{(1)} (\nu-i\delta ) &=& i \pi \rho n_{b} + \rho \ln \frac{D}{
\nu-i\delta }\cr
\rho {\cal  J} (\nu-i \delta  )&=& \frac{\rho }{\frac{1}{J_{0}}- \Pi
(\nu-i\delta )} 
= - \frac{1 }{\ln \left(\frac{T_{K}
}{|\nu| } \right)+i \pi [n_{b}+\theta (-\nu)]}, 
\end{eqnarray}
corresponding to an effective Kondo interaction ${\cal J }(\nu)$
which changes sign, from positive and 
antiferromagnetic ($Re[{\cal }J]>0$) 
at high energies to negative and ferromagnetic ($Re[{\cal }J]<0$)
below  $\nu \sim T_{K}$.  
Since the interactions become weak at both low and high energy, this
expression captures the essential character of the full solution in
these limits.  To fine-tune the solution, we do however need to take
account of the renormalization of the local conduction electron
density of states at low energy. 
The renormalized conduction electron
propagator at the Fermi energy is given by
\[
g (0-i\delta ) = \frac{g_{0}}{1- g_{0}\Sigma (0 - i \delta )} = 
\frac{i\pi\rho  }{1 - i \pi\rho \Sigma (0 - i\delta )}
\]
Now we can determine $\Sigma (0)$ by noting that at the Fermi energy,
the scattering is elastic, so that $\Sigma (0)$ is real. From the
conduction electron phase shift, 
\[
\delta _{c}= \pi k = {\rm  Im }\ln \left[1 - i \pi \rho 
\Sigma (0 - i \delta ) \right],
\]
it then follows that
\[
\Sigma (0)= - \frac{\tan (\pi k)}{\pi \rho }
\]
from which we deduce that
\begin{eqnarray}\label{l}
1-i \pi \rho \Sigma (0- i\delta ) &=& e^{ i \pi k } \sec
(\pi k ), \cr
g (0 - i \delta )&=& i \pi\rho e^{-i \pi k } \cos \pi k 
\end{eqnarray}
so the renormalized density of states is given by $\rho^{*}=
\frac{1}{\pi}Im [g (0 - i\delta )]= \rho \cos^2(\pi k
)$.  This {\sl depression} in the local density of states will reduce
the coefficient of $\ln (\nu)$ in the frequency dependence of the
inverse coupling constant. If 
we approximating the renormalized density of states by $\rho^{*} $
within an energy $T_{K}$ of the Fermi energy and $\rho $ otherwise, 
we may write an improved approximation for $\Pi (\nu)$, as 
\begin{eqnarray}\label{}
\Pi^{(2)} (\nu-i\delta ) &=& \rho^{*}\left( i \pi n_{b} +  \ln \frac{T_{K}}{
\nu-i\delta }
\right)+ \rho\ln \frac{D}{T_{K}}
\cr
\rho {\cal  J} (\nu-i \delta  )&=& \frac{\rho }{\frac{1}{J_{0}}- \Pi
(\nu-i\delta )} 
= - \frac{\sec^{2} (\pi k ) }{\ln \left(\frac{T_{K}
}{|\nu| } \right)+i \pi [n_{b}+\theta (-\nu)]}, 
\end{eqnarray}
The logarithmic scaling and the dependence  $\frac{1}{\rho {\cal
J}}\sim \cos^{2} (\pi k ) \ln  (T_{K}/\vert \nu\vert )$ are
indeed confirmed in our numerical results (Fig. \ \ref{scalingplot}).
One of the interesting points about this result, is that the
propagator is logarithmically dependent on energy and does not develop
a power-law dependence on energy that is the hallmark of the overscreened
Kondo model\cite{parcollet97b}, and also appears to be present in a
recent fermionic parallel to the current approach\cite{florens}.
The  $\chi $ fermion basically
represents the singlet  combination of conduction electron and
Schwinger boson $\chi_{\mu}\sim \sum_{\sigma }\psi \dg
_{\sigma}b_{\sigma \mu}$.  
The absence of a powerlaw in the $\chi $ propagator
is evidence that the addition, or removal of a boson from the system
does not change the scattering phase shift. This is presumeably because
the addition or removal of bosons from the system merely changes the
size of the undescreened moment, without altering the number of bosons
that are bound into the singlet. 
\figwidth=12cm
\fg{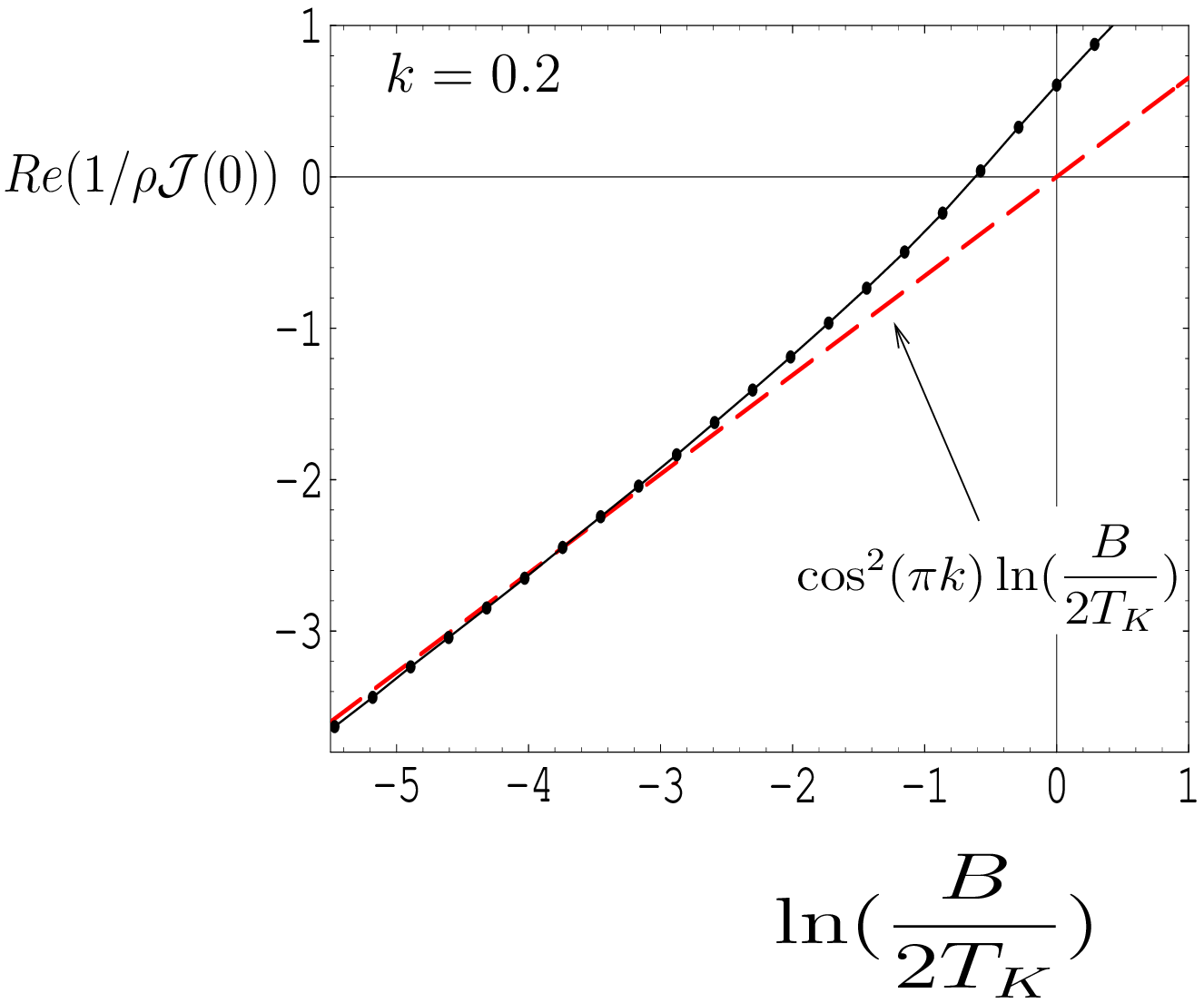}{scalingplot}{Showing 
logarithmic dependence of $Re[1/{\cal
J} (0)]$ on magnetic field at zero temperature, calculated numerically
for the case $k=0.2$. The magnetic field
provides the cut-off to the logarithmic scaling.  For details of how
the magnetic field was introduced, see section VI. Dashed line is the
the curve $y=\cos^{2} (\pi k)  ln (\frac{B}{2T_{K}})$. 
}

At low temperatures ${\cal J} $ scales to zero with logarithmic
slowness. If we look at the the conduction electron self-energy as
given by (\ref{core1}), we
see that the leading singular frequency dependence is given by the
$k n_b{\cal  J} (\omega)$ term, so that at low frequencies,
\begin{eqnarray}\label{l}
\Sigma (\omega) = - \frac{k ( n_{b}+ \theta (\nu )) \sec^{2} (\pi k)}{
\rho \left[ 
\ln \left(\frac{T_{K}
}{|\omega| } \right)+i \pi [n_{b}+\theta (\omega)]
\right]} - \frac{\tan (\pi k)}{\pi \rho }
=- \frac{\tan (\delta (\omega))}{\pi \rho }
\end{eqnarray}
where 
\begin{eqnarray}\label{phaseshift2}
\delta _{c} (\omega)\sim 
\pi k\left[ 1+ \frac{ [n_{b}+\theta (\omega)] }{\ln
(T_{K}/|\omega|)}\right] 
+O\left(\frac{1}{\ln ^{2} (T_{K}/\vert \omega\vert )} \right).
\end{eqnarray}
This singular energy dependence of the conduction electron phase shift
is a consequence of coupling between
the degenerate manifold of the partially screened moment and the
conduction sea. 
Although the scattering of conduction electrons is elastic at low
energies, the  singular frequency dependence displayed here sets this
system apart from a conventional Landau Fermi liquid (where the phase
shift depends linearly on energy). 
Indeed, if we use
$\Sigma (\omega)$ to define a frequency dependent wave-function
renormalization constant, we find that this quantity diverges as
$Z = \frac{1}{1- \Sigma' (\omega)}\sim
\frac{1}{\omega \log (T_{K}/\omega)}$, so that there are no
well-defined quasiparticles associated with the Kondo scattering. This
is the meaning of the term
``singular'' Fermi liquid. 

The singular energy dependence of the scattering is also reflected in the 
shape of the Abrikosov Suhl resonance, given by  the imaginary part of
the electron t-matrix
\begin{equation}\label{}
Im [t (\omega)] = Im \left[\frac{\Sigma (\omega)}{1 - g_{0} (\omega) 
\Sigma (\omega)} \right]
\sim  
\frac{1}{\pi \rho } 
\left[ \sin^{2} (\pi k)+ \frac{
\pi k [n_b+\theta (\omega)]\sin (2\pi k)}
{\ln (T_{K}/|\omega|)}
  \right].
\end{equation}
These basic features are each  borne out in the detailed numerical solution
of the integral
mean-field equations of the large $N$ limit.  
Fig. \ref{fig2x}.  shows the results of this numerical calculation,
showing the spectral function of the t-matrix  
and the dependence of the phase shift on the number of flavors. 
\figwidth=16cm
\fg{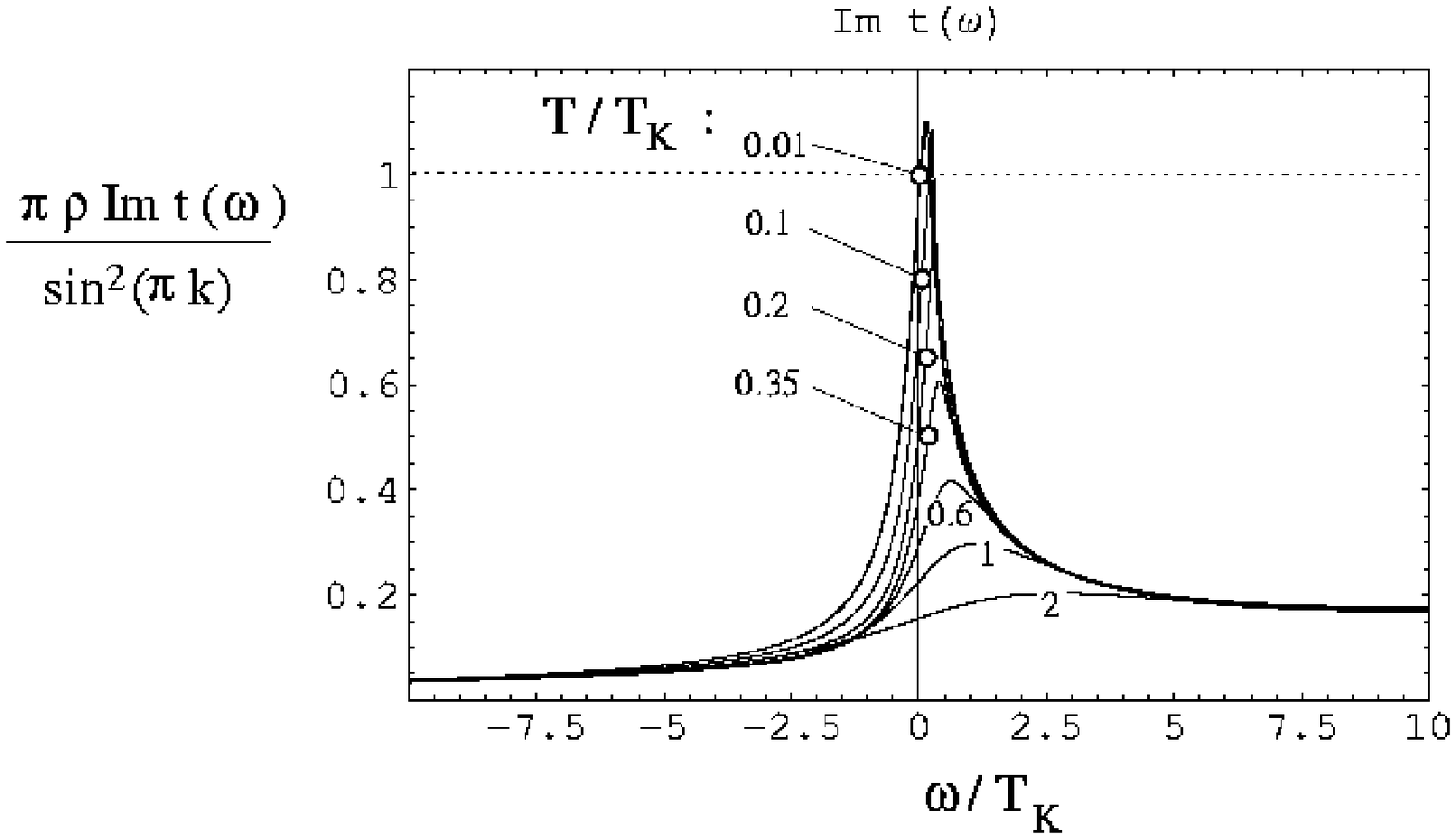}{fig2x}{Showing the frequency dependence of the
t-matrix, normalized with respect to zero temperature value at Fermi
energy.  In this numerical calculation, $k=K/N=0.45$, $n_{b}= 2S/N =
0.2$. Numerical labels give temperature $T/T_{K}$. 
}

This singular energy dependence of the t-matrix also manifests itself
in the temperature dependent resistivity, given by 
\begin{equation}\label{}
\frac{\rho (T)}{\rho _{U}} = n_{i} \pi \rho  \int d\omega
\left(-\frac{\partial f}{\partial \omega} \right) Im t (\omega)
\end{equation}
where $\rho _{U} = \left(\frac{ne^{2}}{m}\frac{2}{\pi \rho }
\right)^{-1}$ is the unitary resistivity as shown in
Fig. \ref{fig2z}.
\figwidth=14cm
\fg{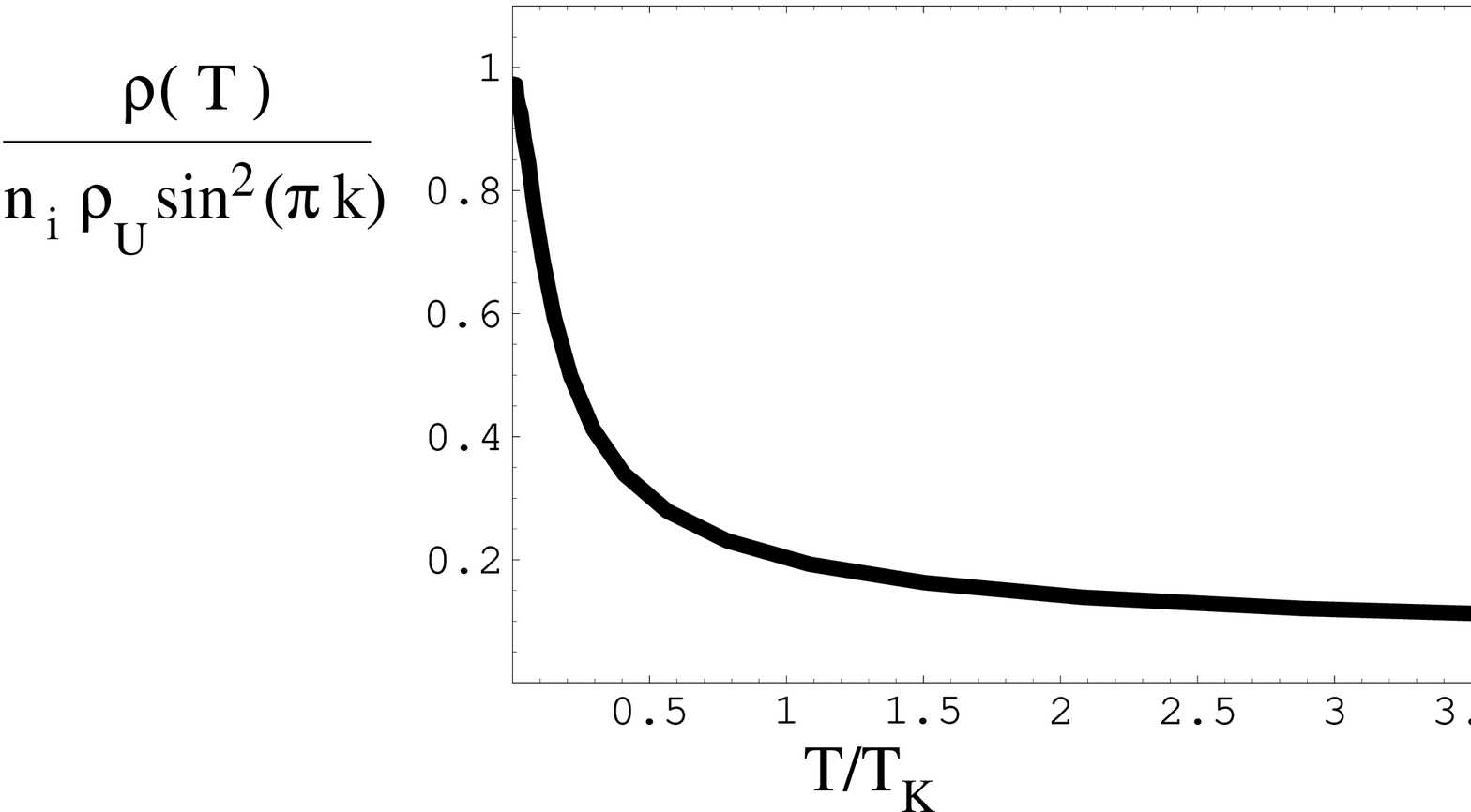}{fig2z}{Showing the temperature dependence of
the single impurity resistance $\rho (T)$, per unit concentration of impurity. 
For this plot, $k=K/N=0.45$, $n_{b}= 2S/N = 0.2$. 
}
\figwidth=10cm
\section{Phase shift and the screening of the moment}\label{}

To reveal the screening of the local moment, we need to compute the 
first correction to the constraint equation.
To satisfy the constraint  we must differentiate the Free energy (\ref{freeenergy})
w.r.t $\lambda$. This yields 
\begin{eqnarray}\label{}
2SK &=&  K   N n_{b} + \delta {\cal N}_{b}\cr
\delta {\cal N}_{b}&=&
- T^{2}{K}\sum_{i\nu_{n}, i\omega_{r} } 
\frac{1}{(i \nu_{n}-\lambda)^{2}}
 g (i \nu_{n}+i \omega_{r}){\cal  J} (i \omega_{r}).
\end{eqnarray}
The first term in this sum determines the number of bosons that
condense into the unscreened magnetic moment. The second term
$\delta {\cal N}_{b} $ is the total number of bosons bound into the Kondo
singlet in the ground-state.  

At zero temperature we can replace the
discrete Matsubara sums by a continuous integral as follows,
\[
T\sum_{i\nu_{n} } \longrightarrow
\int_{-i\infty }^{{i\infty }} 
\frac{d\nu }{2 \pi i},\qquad T\sum_{i\omega_{r} } \longrightarrow
\int_{-i\infty }^{{i\infty }} 
\frac{d\omega }{2 \pi i},
\]
so that the number of bound-bosons becomes
\begin{eqnarray}\label{ps1}
\delta {\cal  N}_{b} = 
K \int \frac{d \omega }{(2 \pi i)}
\int \frac{d \nu }{(2 \pi i)}
g ( \nu + \omega) \frac{\partial }{d\nu} \left(
\frac{1}{\nu-
\lambda} \right) {\cal  J} (\omega), 
\end{eqnarray}
where we have replaced $\frac{1}{( \nu-\lambda)^{2}}
\rightarrow -\frac{\partial }{d\nu}
\left(
\frac{1}{( \nu-\lambda) }
\right)
$.  Next we can integrate the internal integral by parts
\begin{eqnarray}\label{ps1b}
\delta {\cal  N}_{b} = 
-K \int \frac{d \omega }{(2 \pi i)}
\int \frac{d \nu }{(2 \pi i)}
\frac{\partial }{d\nu} 
\left(g ( \nu + \omega) \right) 
\frac{1}{\nu-
\lambda} 
{\cal  J} (\omega), 
\end{eqnarray}
and then replace $\frac{\partial }{d\nu} 
\rightarrow \frac{\partial }{d\omega}  
$ 
to obtain
\begin{eqnarray}\label{ps1c}
\delta {\cal  N}_{b} = 
K \int \frac{d \omega }{(2 \pi i)}
\frac{\partial }{d\omega} 
\overbrace {\left(-\int \frac{d \nu }{(2 \pi i)}
g ( \nu + \omega) 
\frac{1}{\nu-
\lambda} \right) }^{{\Pi (\omega)}}
{\cal  J} (\omega), 
\end{eqnarray}
We can now identify the term inside the central brackets as the
self-energy $\Pi (\omega)$, which enables us to compactly rewrite the
integral as 
\begin{eqnarray}\label{}
\delta {\cal  N}_{b} &=& 
K \int \frac{d \omega }{(2 \pi i)}
\frac{\partial }{d\omega} 
\left(\Pi (\omega)\right) {\cal  J}
(\omega)\cr
&=& -K \int \frac{d \omega }{(2 \pi i)} 
\frac{\partial }{d\omega} 
\ln \left[- {\cal J}^{-1}
(\omega) \right] \cr
&=& -\frac{K}{\pi}  \left[
\delta _{\phi } (0)- \delta _{\phi } (-\infty
) \right] = -\frac{K}{\pi} \delta _{\phi } 
\end{eqnarray}
where
\[
\delta _{\phi } (\omega) = {\rm  Im }\ln \left[
\frac{1}{J}- \Pi(\omega-i \delta ) 
\right].
\]
so that 
\[
K 2S= \overbrace {NK n_{b}}^{2K S^{*}} -\overbrace {\frac{K}{\pi}\delta _{\phi }}^{{\delta n_{b}}}.
\]
Now the argument inside the logarithm that determines $\delta _{\phi}$
is the inverse of the running coupling constant, 
\[
\frac{1}{{\cal J} (\omega)}
=\frac{1}{J}- \Pi(\omega-i \delta ).
\]
In the underscreened Kondo model, the residual coupling between the
partially screened local moment and the conduction electrons  becomes
ferromagnetic at low temperatures, scaling logarithmically to zero.
This means that $\frac{1}{{\cal J} (\omega-i\delta )}\sim - \rho ^{-1}\ln
(T_{K}/\omega)- i \delta $ at low energies, which implies that $\delta
_{\phi }=-\pi$. This in turn, implies that 
the number of bound-bosons is $\delta {\cal N}_{b}=K$. 
As a consequence the number of unpaired bosons is reduced by $K$
in the ground-state, and if we define $S^{*}= \frac{N}{2} n_{b}$, the
effective spin of each flavor will be
\[
S^{*} = S - \frac{1}{2}.
\]
We can also relate the number of bound bosons to the conduction
electron phase shift $\delta _{c}$, in a parallel fashion, as follows:
\begin{eqnarray}\label{ps2}
\delta {\cal  N}_{b}&=& K \int \frac{d \omega }{(2 \pi i)}
\frac{\partial }{d\omega} 
\overbrace {\left(\int \frac{d\nu}{2 \pi i}
{\cal  J} (\omega- \nu )
\frac{1}{\nu-
\lambda} \right)}^{-\Sigma (\omega)} g ( \omega) 
\cr
&=& -N \int \frac{d \omega }{(2 \pi i)}
\frac{\partial \Sigma (\omega)}{d\omega} 
{g}
(\omega)\cr
&=& N \int \frac{d \omega }{(2 \pi i)} 
\frac{\partial }{d\omega} 
\ln \left[g_{o}^{-1}- \Sigma(\omega) \right] \cr
&=& \frac{N}{\pi}  \left[\delta _{c} (0)- \delta _{c} (-\infty
) \right] = \frac{N}{\pi} \delta _{c} .
\end{eqnarray}
In deriving the third line of this expression, we have taken the large
band-width limit, enabling  all frequency dependence of 
the conduction green function $g_{o}= - i \pi\rho {\rm sign }
(\omega_{n})$ to be be ignored. 
 The conduction electron phase shift
is given by
\begin{equation}\label{}
\delta _{c } (\omega) = {\rm  Im }\ln \left[
1- g_{0} (\omega-i\delta )\Sigma(\omega-i \delta ) 
\right].
\end{equation}
By comparing the two expressions (\ref{ps1} ) and (\ref{ps2} ) 
for $\delta {\cal N}_{b}$, we are able to confirm 
that the conduction electron phase shift is given by
\[
\delta _{c}= \pi 
\left(\frac{\delta {\cal  N}_{b}}{N}\right)
= \pi k.
\]
Notice that along the way we have proven that
\begin{equation}\label{}
N 
\left(\frac{\delta _{c}}{\pi }  \right)
+ K \left(\frac{\delta _{\phi }}{\pi}  \right)
=0
\end{equation}
Although we have proven this strictly in the large N limit, this
result is a Ward Identity that is expected to hold for all $N$, 
a result which 
relies solely on Fermion number conservation (see Appendix A). This result is in
effect,  a statement of famous the ``Anderson-Clogston compensation theorem'' - 
that the total number of bound fermions bound by the Kondo effect in
the infinite band-width limit, is zero\cite{clogston61}. 

\section{Zero temperature Magnetization}\label{}

In order to examine the effect of a magnet field  in the large $N$
limit we need to be careful about our definition of the magnetization.
We shall suppose that the magnetic field couples preferentially
to the ``up'' spin (s) of each flavor, i.e. that the magnetization takes
the form
\[
\hat M = g_{N}(\hat n_{b\uparrow\mu} -\hat
n_{b\uparrow\mu}\vert _{B=0} )
\]
There are various ways in which we can now definite $\hat  n_{b\uparrow
\mu}$.  One way to do this, is to define the ``up'' states as the
first $K$ spin components, i.e
\[
\hat M^{(1)} = \frac{N}{2 (N-K)} \left[\sum_{\mu, \si =1}^{K} \hat n_{b\si\mu} - \frac{2SK^{2}}{N}
 \right]
\]
Here, value of $g_{N}$ is chosen so that at full polarization,
$M=SK$.  With this definition, the first $K$ spin channels  of the
conduction electron are ``up'' electrons 
and the remaining 
$N-K$ spin channels are ``down''.  When we add $\hat M^{(1)}$ to the
corresponding expression for the conduction electrons, we form a
conserved quantity, so that we shall call $M^{(1)}$ the ``conserved
magnetization''. 
By imposing the constraint, $\sum_{\sigma\mu }n_{b\sigma
\mu}= 2SK$, 
the conserved magnetization can be rewritten as
\[
\hat M^{(1)} = \frac{1}{2 }\left[2SK - \frac{N}{N-K} n_{b\downarrow }
\right]
\]
where $n_{b\downarrow }= 2SK - n_{b\uparrow}= \sum _{\mu=1}^{K}
\sum_{\sigma >K}n_{b\sigma \mu} 
$.
Now unfortunately, if we couple the magnetic field up to $M^{(1)}$,
then we do not completely remove the spin degeneracy of the
ground-state.  For this purpose, we need a more restrictive definition
of $n_{b\uparrow}$, and we shall choose ``up'' to mean the spin
component where $\mu = \sigma $, i.e 
\[
n_{b\uparrow}\equiv  \sum_{\mu=1}^{K}\sum_{\sigma =1}^{N} n_{b\sigma\mu}\delta _{\mu
\sigma }= \sum_{\mu=1}^{K}n_{b\mu \mu}.
\]
With this definition, by imposing the constraint, $\sum_{\sigma\mu }n_{b\sigma
\mu}= 2SK$, we obtain 
\[
\hat  M = K S - \frac{1}{2}\sum_{ \sigma \ne a}n_{b\sigma a}
\]
We shall use this definition as our method for
coupling the magnetic field to the spin, so that $M= - \delta F/\delta
H$ is the true ``thermodynamic'' magnetization. 
In the case where $K=1$, the thermodynamic and conserved
magnetizations correspond exactly. However, the need to break the
flavor or replica
symmetry at finite $K$ unfortunately  forces us to delineate between these two forms.
However, we shall see shortly that $\langle M^{( 1)} \rangle  $ 
and $\langle M \rangle$ 
have almost identical expectation values, and that the
conserved magnetization has a far more convenient expression in terms
of the scattering phase shift. 
The coupling to a magnetic field then gives the Hamiltonian
\begin{eqnarray}\label{}
H_{B}=  -B\hat  M = -B KS +\frac{B}{2}\sum_{\sigma \ne a
}n_{b\sigma a}.
\end{eqnarray}
Now in the large $N$ limit at low temperatures in a field, the ``up''  bosons
condense. This allows us to carry out the Kondo version of spin-wave
theory. Formally, we condense the ``up'' bosons and integrate over their
phase fluctuations to exactly impose the constraint.  The resulting
Holstein Primakoff transformation is obtained by replacing
\begin{equation}\label{}
\hat b_{\sigma a} \rightarrow \left\{
\begin{array}{cr}
\sqrt{2S-n_{ba}}&(\sigma=a ),\cr
b_{\sigma a}&(\sigma \ne a)
\end{array} \right.
\end{equation}
where $n_{ba}=\sum_{\sigma \ne a}
b\dg _{\sigma a}
b_{\sigma a}$. In this process, 
the $\sigma =a$ or ``up'' components of the boson fields have 
been eliminated. 
In a field, the amount of fluctuations $n_{ba}\sim O (1)$, so that as
in spin-wave theory, to
leading order, we can drop the $n_{ba}$ inside the square-root. With this
understanding, inside the interaction, we must now make the replacement
\[
\sum_{\beta }\psi \dg _{\beta }b_{\beta a}\rightarrow \psi \dg
_{a}\sqrt{2S}+ \sum_{\beta' \ne a}\psi \dg _{\beta '}b_{\beta 'a}
\]
so that the Hamiltonian in a field now becomes
\begin{eqnarray}\label{H1}
H&=& \sum_{k\alpha } \epsilon _{k}c\dg _{k\alpha }c_{k\alpha
}+H_{I}+H_{B}\cr
H_{I}&=& \frac{1}{\sqrt{N}}\left\{ \sum_{a=1}^{K}\left( 
\psi \dg
_{a}\sqrt{2S}+ \sum_{\beta' \ne a}\psi \dg _{\beta '}b_{\beta 'a}
\right)\phi _{a} + H.c \right\}
-\sum_{a=1}^{K}\frac{\bar {\phi }_{a}\phi _{a}}{J}\cr
H_{B}&=& B\left(\frac{1}{2}\sum_{a=1}^{K}n_{ba}- KS \right)
\end{eqnarray}

The effective action in a field now becomes
\begin{eqnarray}\label{}
S &=& \int _{0}^{\beta }d\tau  
\left[ \sum_{k,\sigma }c\dg
_{k\si}\left(\partial_{\tau }+\epsilon_{k} \right)c_{k\si} -\frac{1}{J}
\bar \phi_{\mu} \phi_{\mu} - SK
(2\lambda +B)
\right]+S_{I}+S_{mix}\cr
S_{I}&=& -\frac{1}{N}\sum_{a=1}^{K}\sum_{\sigma \ne a}
\int_{0}^{\beta } d\tau d\tau '
 \psi \dg _{\sigma } (\tau )\psi _{\sigma } (\tau' )
D_{\sigma } (\tau -\tau ')
\bar \phi _{a} (\tau ') \phi_{a} (\tau )
\cr
S_{mix}&=& \sqrt{\frac{2S}{N}}\sum_{a=1}^{K}\int_{0}^{\beta } d\tau
\left[\psi (\tau )\dg _{a}\phi (\tau )
_{a}+ \bar \phi _{a} (\tau )\psi _{a} (\tau ) \right]
\end{eqnarray}
where $D_{\sigma } (i\nu_{n})= \frac{1}{i \nu_{n}- (B/2)}$.
The interaction term $S_{I}$ can be split up into a term with no restrictions
on the spin flavor summations, plus a term  ${\cal  R}$ that can be neglected in
the large $N$ limit:
\begin{eqnarray}\label{}
S_{I}&=& -\frac{1}{N}\sum_{a=1}^{K}\sum_{\sigma =1}^{N}
\int_{0}^{\beta } d\tau d\tau '
 \psi \dg _{\sigma } (\tau )\psi _{\sigma } (\tau' )
D_{\sigma } (\tau -\tau ')
\bar \phi _{a} (\tau ') \phi_{a} (\tau )+{\cal R}\cr
\cal R&=&\frac{1}{N}\sum_{a=1}^{K}
\int_{0}^{\beta } d\tau d\tau '
 \psi \dg _{a } (\tau )\psi _{a } (\tau' )
D_{\sigma } (\tau -\tau ')
\bar \phi _{a} (\tau ') \phi_{a} (\tau ).
\end{eqnarray}
If we carry out a Hubbard Stratonovich decoupling on the first term, we
now have:
\begin{eqnarray}\label{}
S  &=& 
\int_{0}^{\beta } d\tau d\tau ' {\cal L} (\tau ,\tau ') - \beta S K B\cr
{\cal L}&=&
\sum_{\sigma>K }
\psi \dg_{\sigma }\left[-g_{0}^{-1}+\Sigma  
\right]
\psi_{\sigma }  
+ 
\sum_{a=1}^{K} \left(
\psi \dg _{a},\bar \phi _{a} 
\right)
\left[
\begin{matrix}
-g_{0}^{-1}+\Sigma  
& \sqrt{\tilde{n}_{b}}
\cr
\sqrt{\tilde{n}_{b}}& -J^{-1}+
\Pi 
\end{matrix}
\right]
\left(\begin{matrix}
\psi _{a}\cr
\phi _{a}
\end{matrix}
 \right)
\\ &+& 
N \Pi (\tau -\tau ') \frac{1}{D_{0} (\tau -\tau ')} \Sigma (\tau '-\tau )
\end{eqnarray}
where $\tilde{n}_{b}= 2S/N$
Taking the saddle point values for  $\Pi$ and $\Sigma$, 
and 
then 
integrating out the fermions, the mean-field free energy in a field is
then 
\begin{eqnarray}\label{}
\frac{F}{N T} &=&- (1-k)  {\rm  Tr}\ln \left[- g_{o}^{-1}+\Sigma
\right] - k {\rm  Tr}\ln
\left[
\begin{matrix}
-g_{0}^{-1}+\Sigma  
& \sqrt{\tilde{n}_{b}}
\cr
\sqrt{\tilde{n}_{b}}& -J^{-1}+
\Pi 
\end{matrix}
\right]\cr
&+& N\int d\tau d\tau '
\Pi (\tau ',\tau )\frac{1}{D_{0} (\tau -\tau ')}\Sigma (\tau ,\tau ')
- SK B.
\end{eqnarray}
When we impose the saddle-point condition on $\Sigma$ and $\Pi$, 
decomposition, we must be careful to delineate
between the propagators of ``up'' and ``down'' electrons. The
mean-field equations are then 
\begin{eqnarray}\label{}
\Sigma (\tau -\tau ')&=&  -k D (\tau -\tau ') 
{\cal J} (\tau -\tau ')\cr
\Pi (\tau -\tau ') &=& - D (\tau '-\tau )
[k g_{\uparrow} (\tau -\tau ')+ (1-k)g_{\downarrow }].
\end{eqnarray}
The mixing terms between the ``up'' electrons and the ``phi'' fields
mean that we must now modify the propagators as follows:
\begin{eqnarray}\label{gdownup}
g_{\downarrow} (\omega) &=& \frac{1}{g_{0}^{-1} (\omega)-
\Sigma_{\downarrow} (\omega)}\cr
g_{\uparrow} (\omega) &=& \frac{1}{g_{0 }^{-1}
(\omega)-\Sigma_{\uparrow} (\omega)
}.
\end{eqnarray}
Here
\begin{eqnarray}\label{sigmadwrap}
\Sigma_{\downarrow} (\omega)&=& \Sigma (\omega) 
\cr
\Sigma_{\uparrow} (\omega)&=& \Sigma (\omega) + \tilde{n}_{b}\left(J^{-1} - \Pi (\omega) \right)^{-1}
\end{eqnarray}
and
\begin{eqnarray}\label{jfield}
{\cal J} (\omega) = \frac{1}{J^{-1}- \Pi (\omega) -
\tilde{n_{b}}g_{\downarrow } (\omega)}
\end{eqnarray}
where $\tilde{n}_{b}= 2S/N$.

If we convert the integral equations to Matsubara summations, we
obtain
\begin{eqnarray}\label{kor1}
\Sigma (i \omega_{n}) &=& -k
\sum _{r}\frac{1}{i  \nu_{r} -(B/2)}
 {\cal J} ( i \omega_{n}- i \nu_{r})\cr
\Pi (i \omega_{n}) &=& -\sum _{r}\frac{1}{i  \nu_{r} -
(B/2)} \left( k g_{\uparrow } ( i \omega_{n}+ i \nu_{r})
+ (1-k)g_{\downarrow }( i \omega_{n}+ i \nu_{r})
\right)
\end{eqnarray}

Carrying out the Matsubara summations, we then obtain
\begin{eqnarray}\label{core1x}
\Sigma (z)&=& 
k\int \frac{dy}{\pi}\frac{1- f (y)}{z-y- (B/2)}Im
{\cal J}
(y-i\delta )\\\label{core2x}
\Pi (z)&=& 
\int \frac{dy}{\pi}
\frac{ f (y)}{z-y+ (B/2)}
Im \left[
k g_{\uparrow} (y-i\delta )
+ (1-k)g_{\downarrow} (y-i\delta )
\right]
.
\end{eqnarray}

From the solutions of these equations, we can compute the field
dependent propagators and free energy.  We can
define the following scattering phase shifts associated with the 
conduction electrons and $\phi $ fermion:
\begin{eqnarray}\label{}
\delta _{\phi } (B) &=& {\rm  Im }
\left[ \phantom{\int}\hskip -0.4cm
\ln \bigl(
1- J [\Pi(z )+\tilde{n}_{b}g_{\downarrow } (z)] \bigr )
\right]
_{z=0-i\delta }
.\cr\cr
\delta _{c\sigma  } (B) &=&
{\rm  Im } \left[\phantom{\int}\hskip -0.4cm 
\ln \bigl ( 1- g_{0}\Sigma_{\sigma } (z
)\bigr ) \right]
_{z=0-i\delta },\qquad \qquad (\sigma = \uparrow,\ \downarrow )
\end{eqnarray}
These phase shifts at finite  field are actually
related by Ward identities (see appendix B), and obey the following
relationships at all fields:
\begin{eqnarray}\label{phaseid}
K \delta_{\phi }+ N \delta _{\downarrow } &=& 0\\
K \delta _{\uparrow} + (N-K)\delta _{\downarrow } &=& K \pi
\end{eqnarray}

The differentiation of the Free
energy w.r.t.  field is identical to the differentiation
w.r.t. $\lambda$ carried out in the previous section.  The magnetization
is given by
\begin{eqnarray}\label{}
M &=& SK - \frac{{\delta {\cal N}_{b}}}{2}\cr
\delta {\cal N}_{b}&=&
- T^{2}{K}\sum_{i\nu_{n}, i\omega_{r} } 
\frac{1}{(i \nu_{n}-B/2)^{2}}
 \tilde{g}(i \nu_{n}+i \omega_{r}){\cal  J} (i \omega_{r}),
\end{eqnarray}
where $\tilde{g}= k g_{\uparrow}+ (1-k)g_{\downarrow }$.
If, instead of using the thermodynamic magnetization, we use the
conserved magnetization $M^{(1)}$, then we find that
\begin{eqnarray}\label{}
M^{(1)} &=& SK - \frac{{\delta \tilde{{\cal N}}_{b}}}{2}\cr
\delta \tilde{{\cal N}}_{b}&=&
- T^{2}{K}\sum_{i\nu_{n}, i\omega_{r} } 
\frac{1}{(i \nu_{n}-B/2)^{2}}
 g_{\downarrow }(i \nu_{n}+i \omega_{r}){\cal  J} (i \omega_{r}),
\end{eqnarray}
The difference between the two expressions results from the slight
energies.  The advantage of the second quantity, is that by being
related to a conserved quantity, it is related to the 
scattering phase shifts. Using exactly the same methods
we 
relate $\delta \tilde{{\cal N}}_{b}$ to the $\phi $ and conduction
electron phase shifts, as before, to obtain
\begin{eqnarray}\label{}
M^{(1)} &=& 
KS + \frac{K}{2}\frac{\delta _{\phi } (B)}{\pi}\cr
&=& KS - \frac{N}{2}\frac{\delta _{\downarrow } (B)}{\pi}\cr
&=& K (S-\frac{1}{2}) + \frac{K}{2}\frac{\delta _{\uparrow} (B) -\delta _{\downarrow } (B)}{\pi}
\end{eqnarray}
where we have used the phase-shift identities (\ref{phaseid})
given above. 

We have solved equations (\ref{gdownup}, \ref{sigmadwrap}, \ref{core1x}) at zero temperature, finite
field $B$ by numerical iteration, using
a Fast Fourier transfer routine to carry out the convolutions. 
Fig. \ref{fig3x} shows the  field dependence of the coupling constant
$g (B) = Re[{\cal J} (\omega)\vert _{\omega=0}$, showing the change in
sign of the coupling constant as the system goes from weak coupling at
high fields to strong coupling at low fields.
\figwidth=10cm
\fg{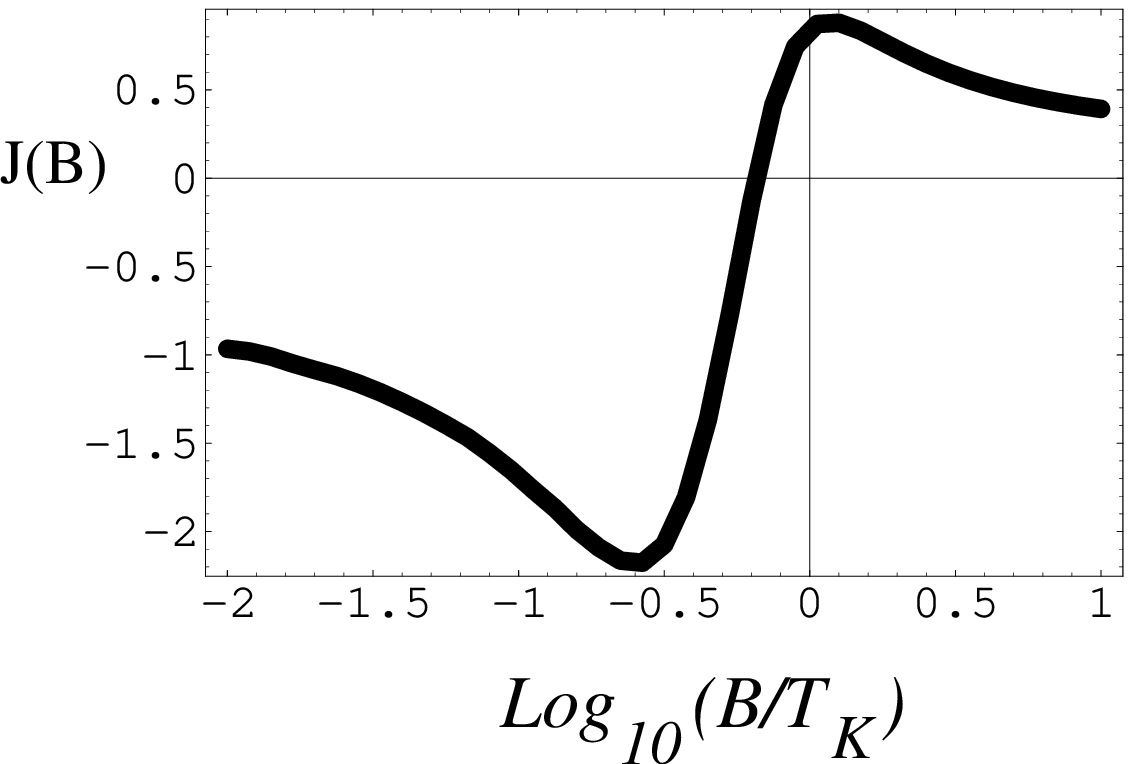}{fig3x}{Dependence of the Kondo coupling constant
on field for  $k=0.45$.}
Fig. \ref{fig3} shows the field dependence of the ``up''  and ``down''
phase shifts. The results obtained by our large $N$ method are
strikingly similar to results recently obtained by numerical
renormalization group and the Bethe ansatz\cite{borda}.
Fig. \ref{fig4} shows the field dependent
reduction in the magnetization $\Delta M= - K \delta _{\phi }
(B)/\pi$. 

\figwidth=12cm
\fg{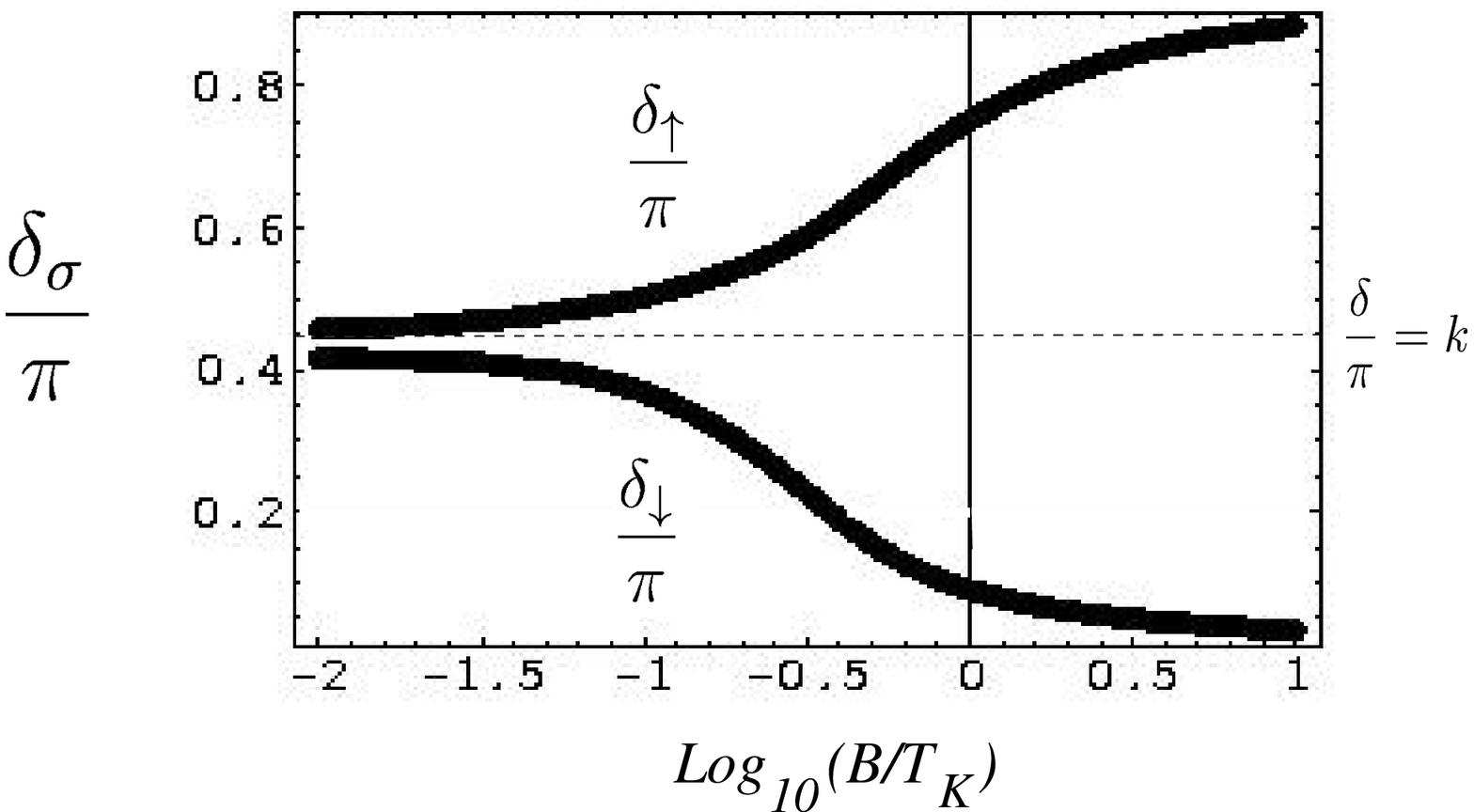}{fig3}{Field dependence of the ``up''  and
``down'' phase shifts for $k=0.45$.}

\figwidth=13cm
\fg{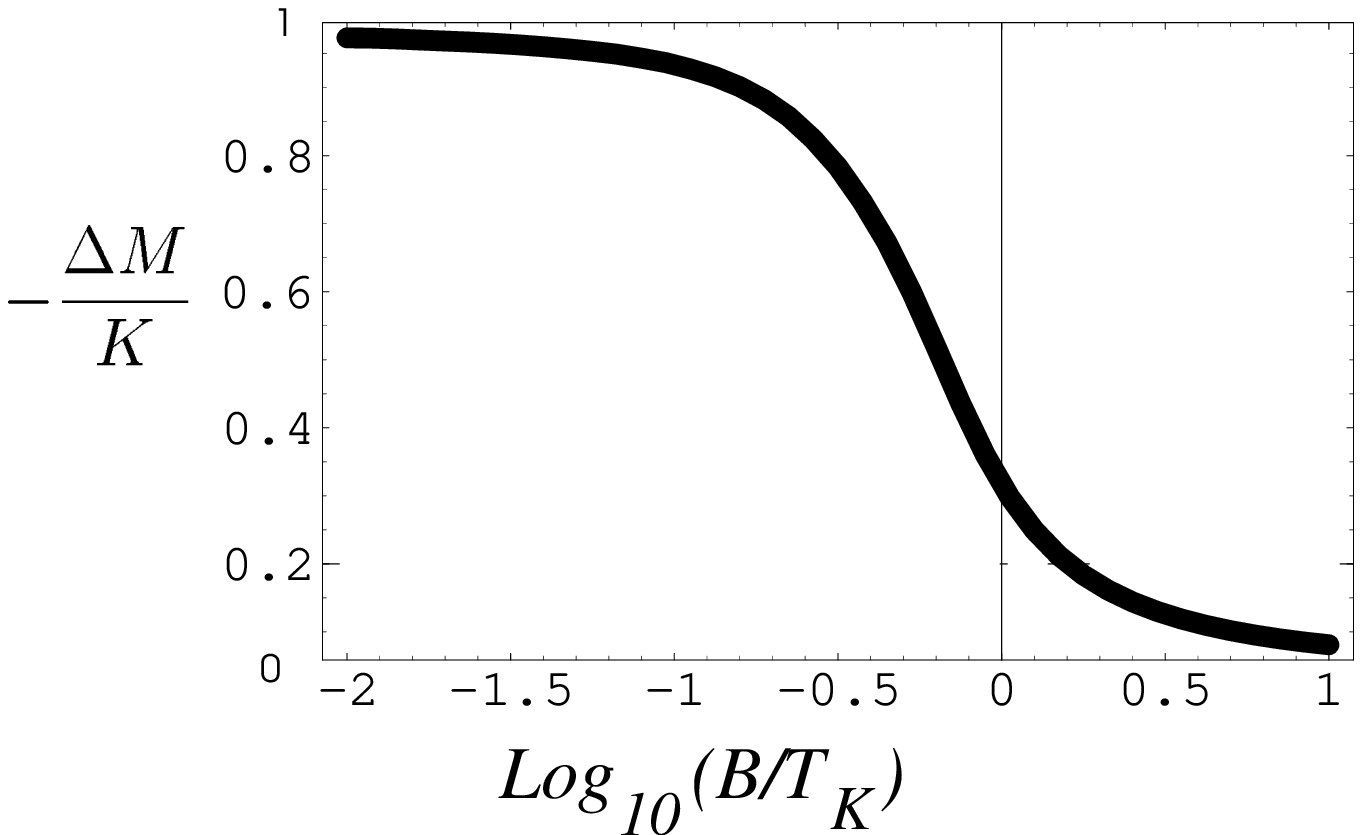}{fig4}{Field dependence of the magnetization reduction $\Delta
M /K = \delta _{\phi }/\pi$
for $k=0.45$
}

\section{Discussion}\label{}

Our method gives a controlled treatment of the underscreened Kondo
model, and using the method of replicas, we have been able to develop
a treatment  of the Kondo effect which correctly reproduces the finite 
phase shift $\delta = \pi K/N$ produced by the scattering resonance. 
\figwidth=8cm
\fgb{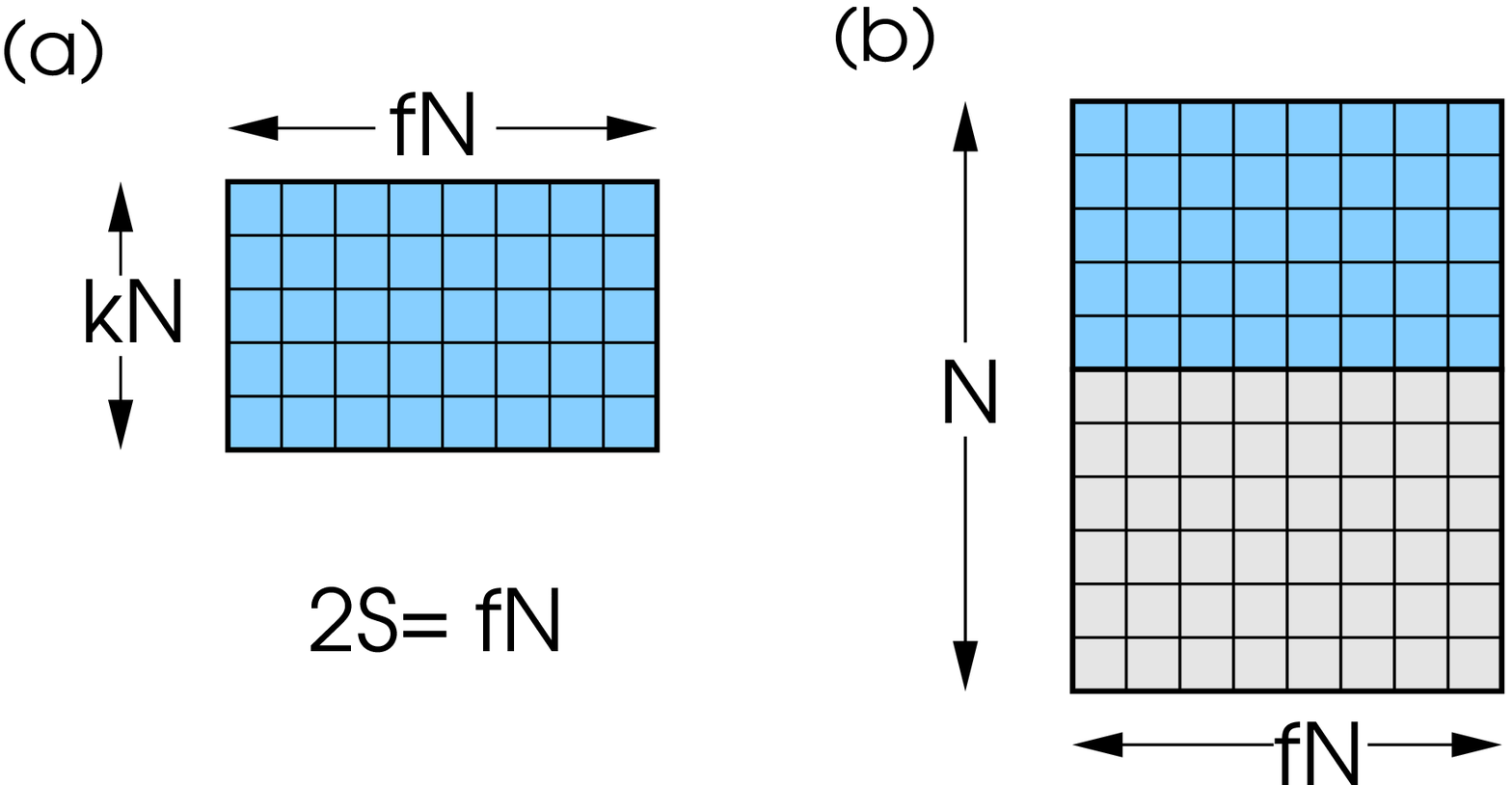}
{fig5}
{(a) Rectangular Young Tableau which will
self-assemble in strong-coupling ground-state (b) of color-flavor-spin
model in the case where $F=2S$.
}
What are the prospects for going further? 

The first question of interest, is whether this approach can be used
to describe the fully quenched Fermi liquid?  An adhoc way to attempt
to access the Fermi liquid, is to seek solutions where 
\[
{\cal  N}_{b}= K = 2KS n_{b} (\lambda)  + N \frac{\langle \delta
_{c}\rangle  }{\pi}
\]
The presence of the additional term in the constraint then raises the
value of $\lambda$, so that in the ground-state $\lambda\ne 0$, and
hence $n_{b}\rightarrow 0$, producing a fully quenched ground-state. 
The obvious drawback with this approach, is that by pushing the problem
to this limit, we are exploring a limit where the occupancy of each boson
spin/flavor state is of order $O (1/N)$, which lies outside the strict region
of validity for the large $N$ expansion.  

A more encouraging approach to the problem may lie in trying to unify the
boson replica approach used here, with the multi-channel approach 
developed by Parcollet and Georges (PG). The PG approach introduces F
flavors of conduction electron, considering an interaction of the form
\begin{eqnarray}\label{HX}
H_{I}&=& 
\frac{J}{N}\sum_{\gamma=1}^{F}\sum_{\alpha \beta  }(\psi \dg _{\beta
\gamma}b_{\beta
})
(b\dg _{\alpha} \psi _{\alpha \gamma}) 
\end{eqnarray}
This approach leads to a scattering phase shift $\delta =\pi/N$ which
vanishes in the large $N$ limit, however, it has the virtues that for $F=2S$
it does describe a fully quenched Fermi liquid, and moreover, the method
can, in the lattice, be neatly combined with the Arovas-Auerbach\cite{arovas}
treatment of antiferromagnetism, using a SP (2N)
description of the RKKY interactions \cite{readsachdev91,zarand}.

Is it possible to extend this model, introducing both conduction
flavor and boson color at the same time? This suggests the following
color-flavor-spin (CFS) model 
\begin{eqnarray}\label{HX}
H_{I}&=& 
\frac{J}{N}\sum_{\alpha ,\beta ,\gamma, \mu}(\psi \dg _{\beta
\gamma}b_{\beta\mu
})
(b\dg _{\alpha \mu} \psi _{\alpha \gamma}) \qquad \qquad (\gamma \in
[1,F], \ \mu \in
[1,K])
\end{eqnarray}
where we choose $2S=F=fN$, and  $K= kN$ in the large $N$ limit. 
The strong-coupling solution to this model involves the formation of a singlet
between a rectangular representation of the Bosons, and the $F$
channels of conduction electron, as shown in Fig. \ref{fig5}.
In the ground-state, we expect the bosons to self-assemble into this
strong-coupling representation. 

To develop a large $N$ expansion of the CFS model, 
we might consider writing down the connected skeleton 
graphs that enter into the Luttinger-Ward\cite{luttingerward} functional  $Y[G,{\cal
J},D]$.
The two leading diagrams in $Y$ take the form shown in Fig. \ref{fig6}. 
\figwidth=10cm
\fg{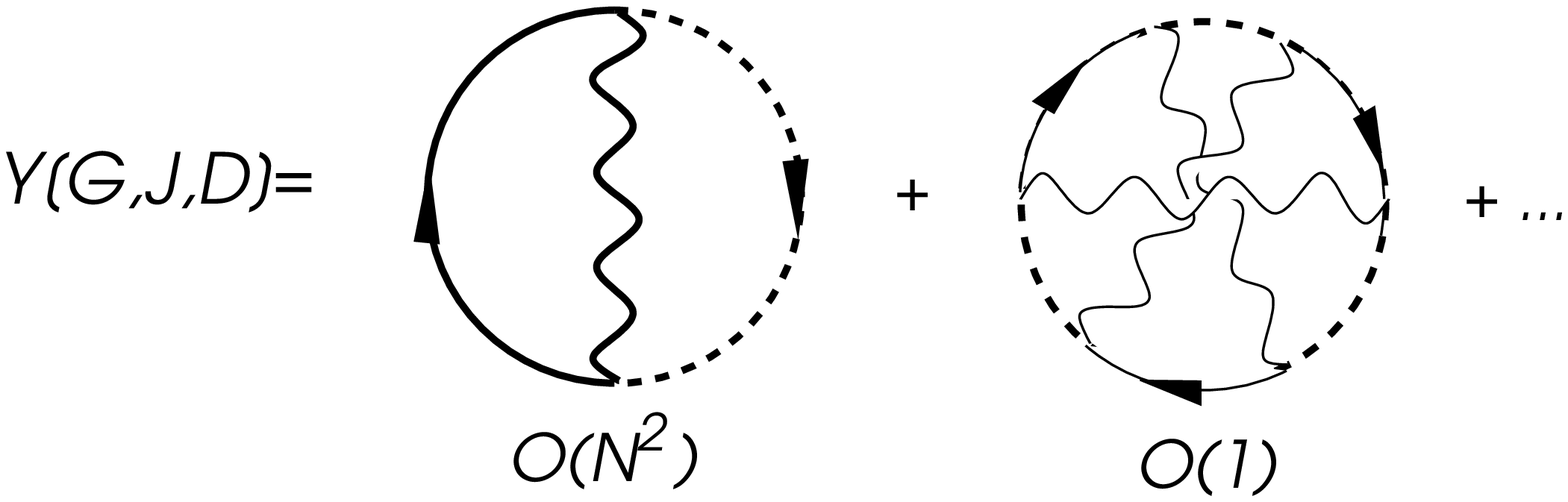}{fig6}{Two leading order skeleton  diagrams for the
Luttinger Ward functional of the CFS model. 
}
There is one loop for each quantum number in each of the above
diagrams, but because the second diagram involves more vertices, it is
smaller by a factor of $O (1/N^{2})$. $Y$ 
can be used to generate the self-energies for a conserving
approximation, via the relationships:
\begin{eqnarray}\label{}
-\frac{\delta Y}{\delta G (i\omega_{n})} &=& NF\Sigma (i \omega_{n}),\cr
-\frac{\delta Y}{\delta {\cal J} (i\omega_{n})} &=& KF\Pi(i \omega_{n}),\cr
\frac{\delta Y}{\delta D(i\nu_{n})} &=& NK\Sigma_{b} (i \nu_{n}),
\end{eqnarray}
where the pre-factors arise because of the sum over spin, flavor and
channel that arises inside $Y$. 
The first term in this series can be used to generate the diagrams that interpolate
between the PG and the replica approach.  This term in the skeleton
expansion of the Free energy is of order $O (N^{2})$ in an approach
that involves color, flavor and spin. Since next skeleton diagram is of
order $O (1)$, so that at first sight, we may neglect all higher order diagrams.

However, it turns out that the development of a strict large $N$
expansion encounters a technical difficulty with the profusion of
higher order planar diagrams. Similar difficulties have been encountered
in the large $N$ treatment of quantum chromodynamics (QCD).  
To see this, it is useful to relabel the conduction, boson and
$\phi$ propagators as two parallel lines, carrying the respective
quantum numbers of spin, color and flavor, as follows
\vskip 0.1truein
\bxwidth=3in
\begin{center}
{\frm{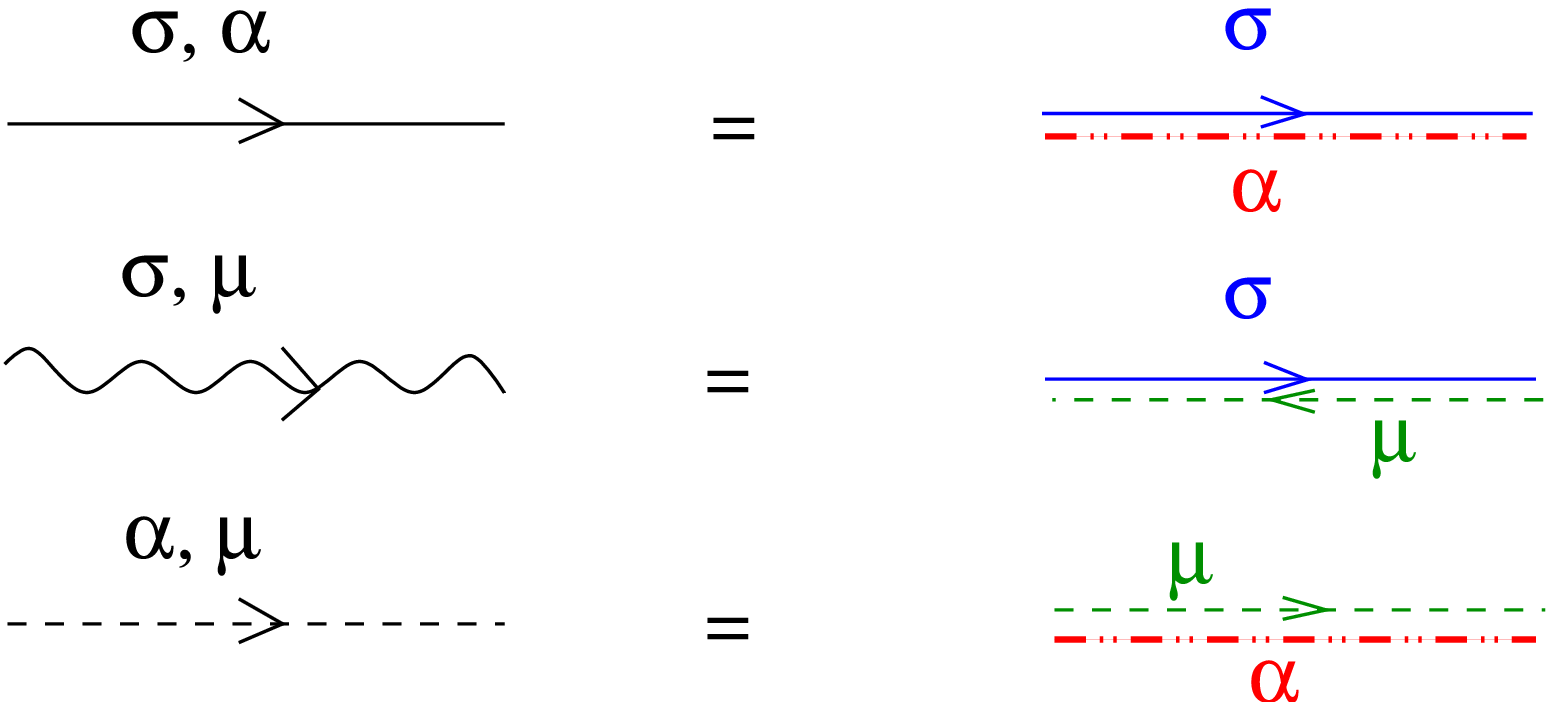}}
\end{center}
\vskip 0.1truein
With this notation, we see that introduction of the additional
flavor quantum number, means that the interaction vertex between
bosons is enhanced by the sum over virtual flavor fluctuations, from 
from a $O (1/N^{2})$ term in the current theories, to a $O (1/N)$
vertex as shown below.
\vskip 0.1truein
\bxwidth=1.5in
\figwidth=1.5in
\fg{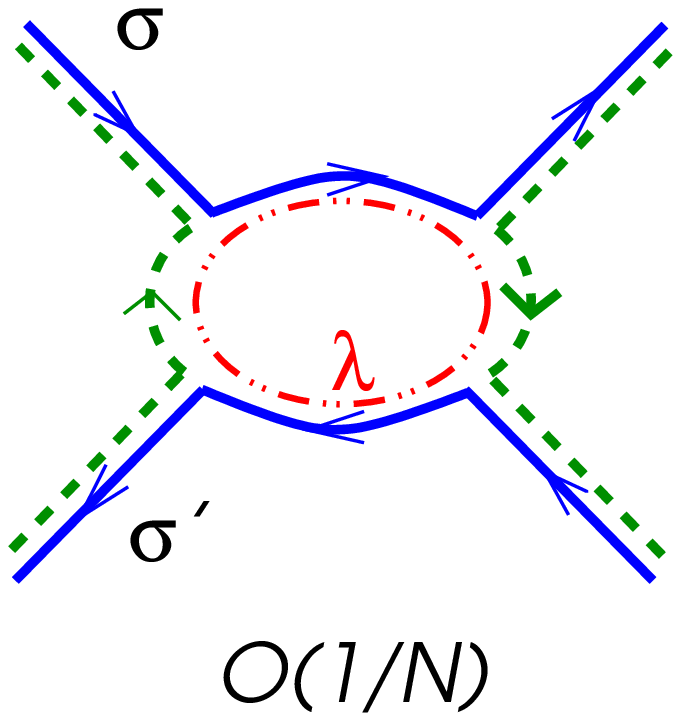}{figloop}{Magnetic vertex.  The additional 
flavor index enhances this vertex by a factor of $N$. }
\noindent If we now look at the boson self-energy diagram
generated by this vertex,
\vskip 0.3truein
\bxwidth=1.5in\upit=-0.35in
\begin{center}
$\raiser{\frm{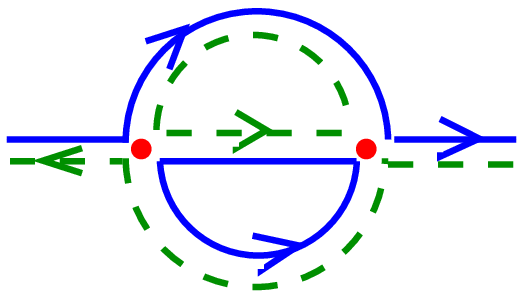}}  =  O (1/N^{2} \times
N^{2} ) =  O (1)$
\end{center}
\vskip 0.2in
\noindent we see it is of order $O (1)$.
Furthermore, when we close the external
legs on this diagram to 
\vfill \eject 
\noindent 
create a skeleton graph, 
\vskip 0.6truein
\bxwidth=2.5in\upit=-0.5in
\begin{center}
$\raiser{\frm{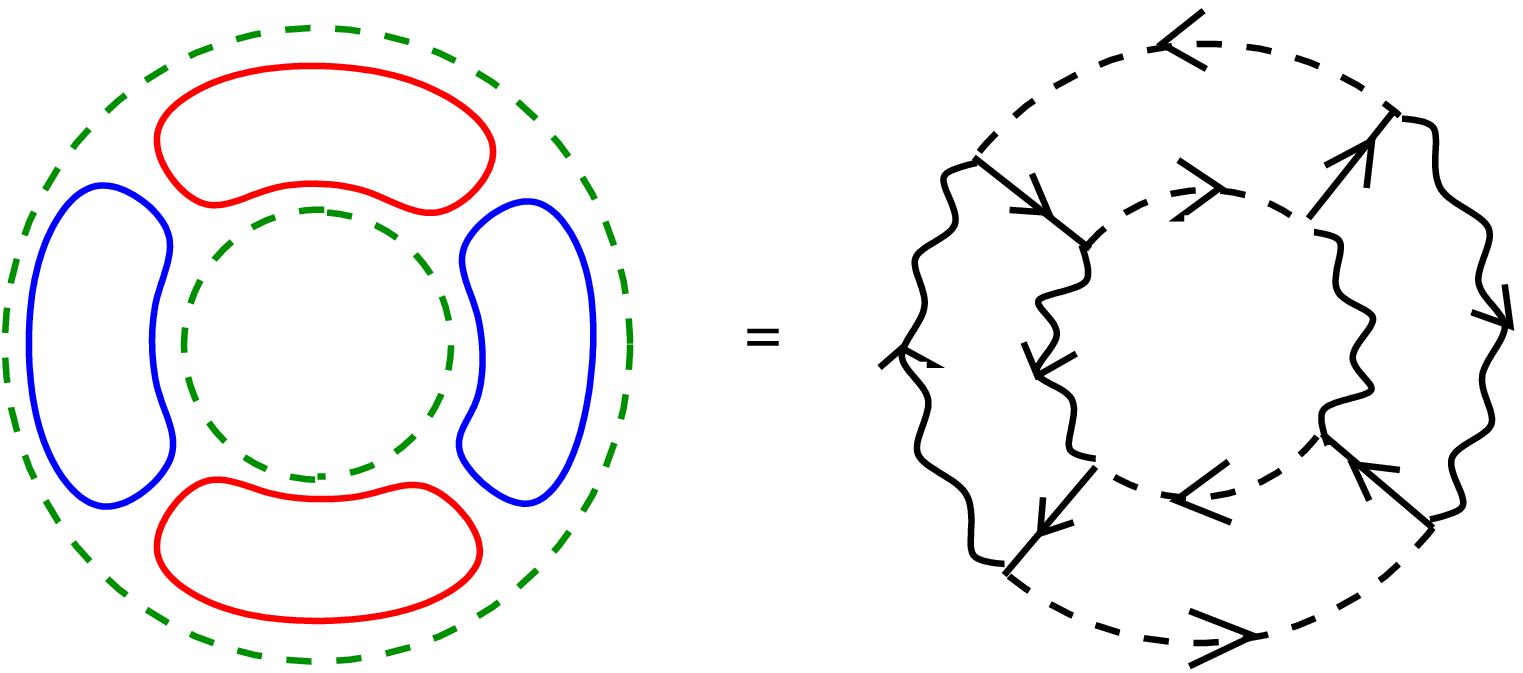}}\qquad = \qquad O (N^{6} / N^{4}
)=O (N^{2})$
\end{center}
\vskip 0.3truein
we see that this graph (which contains six internal quantum number loops,
) is of order $O (N^{2})$- the same order as the leading
graph. Unfortunately, as in QCD, this is just the beginning of 
an entire profusion of higher order
``planar diagrams''  which are all of the same order in the large $N$
expansion.

Despite this difficulty, it may be that the leading order $O (N^{2})$
diagram shown in Fig. 9 is already sufficient to generate a good
conserving mean-field theory for 
the fully quenched Kondo model.  We do not yet know how the neglect of the
higher order planar diagrams affects the results, but it is
tempting to speculate that these diagrams are irrelevant in the
Fermi liquid, or magnetic ground-state, since 
the flavor and color quantum numbers
become massive when the Kondo singlet forms, or when there is an applied
magnetic field. In this case, the higher order diagrams may 
only renormalize the leading order term in the skeleton graph
expansion of the Free energy. These considerations lead us to suggest that the
leading order skeleton free energy diagram for 
the color-flavor-spin model may provide the
key to a successful mean-field theory that spans the magnetic quantum
critical point.

One of the  interesting final questions that deserves discussion,
concerns the physical significance of the additional quantum numbers
that we have introduced. The color-flavor-spin large $N$ approach is basically
approximating the single box  of the Young Tableau representing a spin
$1/2$ by a rectangular Young tableau. What is the meaning of the 
intensive variables $f=F/N$ and $k=K/N$ that appear in the PG and
the current approach? 
One fascinating possibility, is that quantum numbers associated with these
degrees of freedom describe the internal quantum numbers of the
composite quasiparticle in the heavy electron state.  In the Fermi
liquid, we expect these quantum numbers to  be inert, but at the
quantum critical point, these quantum numbers become unconfined. 
These issues will be followed in forthcoming work. 

This research is supported by the
National Science Foundation grant NSF DMR 0312495.  We should
particularly like to thank
Anirvan Sengupta for discussions concerning the matrix aspects of the
CFS model, and the profusion of planar diagrams. Shortly after
submitting this paper to the archive, we became aware of a closely
related fermionic ``replica'' approach by S. Florens,
which we have referenced in this revised draft.\cite{florens}
Discussions with
N. Andrei, G. Kotliar, P. Mehta, G. Zarand and L. Borda  are gratefully acknowledged. 

\section{Appendix A}\label{}

In this appendix, we derive the Ward Identity
\[
N\left (\frac{\delta _{c}}{\pi } 
\right)
 + K \left(\frac{\delta _{\phi }}{\pi } 
 \right)=0
\]
and relate it to the Anderson-Clogston compensation theorem. 
According to this theorem \cite{clogston61}, the net polarization 
of the conduction sea by a localized impurity is zero, in the limit
of a broad band.  A localized resonant scattering center induces Friedel oscillations
in charge density of the medium, and tends to reduce the charge density
in the immediate vicinity of the impurity. The compensation guarantees
that this local depression in charge density is compensated by an
enhancement at greater distances. 

To understand this compensation effect, we need to examine the
conduction electron  Greens function, which is given by
\[
G_{\bf{k},\bf{k}'} (i\omega_{n}) = \frac{\delta _{\bf{k},\bf{k}'}}
{i\omega_{n}-\epsilon_{\bf{k}}} +
\frac{1}
{i\omega_{n}-\epsilon_{\bf{k}}} t (i\omega_{n})
\frac{1}
{i\omega_{n}-\epsilon_{\bf{k}'}} 
\]
where $t (i \omega_{n})$ is the t-matrix of the impurity.  The second
term in this expression induces the Friedel oscillations in charge
density, which are given by
\[
\Delta \rho ({\bf{x}})= NT \sum_{{\bf{k}},{\bf{k}}', i\omega_{n}}\frac{1}
{i\omega_{n}-\epsilon_{\bf{k}}} t (i\omega_{n})
\frac{1}
{i\omega_{n}-\epsilon_{\bf{k}'}}e^{i (\bf{k}-\bf{k}')\cdot
\bf{x}}e^{i\omega_{n}0^{+}}
\]
For example, the change in density at the impurity is given by
\[
\Delta \rho ({\bf{0}})= -N (\pi \rho )^{2} \int_{-\infty }^{\infty
}\frac{d\omega}{\pi} f (\omega){\rm  Im}t (\omega-i\delta ) <0 
\]
( where we have carried out the momentum sum, replacing   $\sum_{\bk }
(i\omega_{n}-\epsilon_{\bk })\rightarrow -i \pi \rho {\rm  sgn} (n)$),
corresponding to a reduction in the electron density. By contrast, the
total change in density, 
\begin{eqnarray}\label{}
\Delta Q = \int d^{3}x\Delta \rho (\bx) = NT \sum_{\bk ,
i\omega_{n}}\frac{1}{(i \omega_{n}-\epsilon_{\bk })^{2}}
t (i\omega_{n})e^{i\omega_{n}0^{+}}
\end{eqnarray}
Assuming that the width of the resonance is much narrower than the
bandwidth $D$, we can replace
\[
\sum_{\bk }\frac{1}{(i \omega_{n}-\epsilon_{\bk })^{2}}
\rightarrow
\rho \int _{-D}^{D}d\epsilon \frac{1}{(i \omega_{n}-\epsilon)^{2}} =
\frac{2 \rho }{D} +O (\frac{\omega_{n}\rho }{D^{2}})
\]
so that 
\begin{eqnarray}\label{}
\Delta Q &=& \frac{2 \rho }{D}T\sum_{i\omega_{n}}t (i\omega_{n})e^{i\omega_{n}0^{+}}
= \frac{\Gamma }{\pi D} \sim 0
\end{eqnarray}
where 
\[
\Gamma = 2 \pi \rho  \int_{-\infty }^{\infty } \frac{d\omega}{\pi} f (\omega){\rm  Im }t
(\omega-i\delta )
\]
is a measure of the width of the resonance. Thus the total
polarization of the conduction band is of order $\Gamma/D<<1$, which
is negligible in the limit of infinite bandwidth. 

Now we shall relate $\Delta Q$ to the scattering phase shifts.  To do
this, we appeal to the Luttinger-Ward functional $Y[G,{\cal J}]$for this
problem, represented by a sum over all skeleton closed loop diagram contributions.
The differential of this functional with respect to the Greens
functions generates the corresponding self-energies:
\begin{eqnarray}\label{}
-\frac{\delta Y}{\delta g_{\sigma } (i\omega_{n})} &=& \Sigma_{\sigma
} (i \omega_{n}),\cr
-\frac{\delta Y}{\delta {\cal J}_{\mu} (i\omega_{n})} &=& \Pi_{\mu}(i \omega_{n}),
\end{eqnarray}
where we have explicitly displayed the spin and flavor indices. 
Now the conservation of charge guarantees that each of these
can be decomposed into one or more closed fermion lines. In the low
temperature limit, the Matsubara sums along these lines may be
replaced by continuous integrals along the imaginary axis, 
\[
T\sum_{i \omega_{n}} \dots \rightarrow \int
\frac{dz}{2\pi i }\dots 
\]
Now if, at zero temperature, the frequency running along each such loop is incremented  by
a small amount $z\rightarrow z+ i \delta \omega$, $Y$ is unchanged,
because the the shift in $z$ can be absorbed by a simple change of
variable. We deduce that at $T=0$, in the absence of a field, 
\begin{eqnarray}\label{}
\delta Y &=& i \delta \omega \int \frac{d\omega}{2 \pi i}
\left[ 
\sum_{\sigma=1 }^{N}
\frac{\delta Y}{\delta g_{\sigma } (\omega)
}\frac{d g_{\sigma }}{d \omega }
+ 
\sum_{\mu =1}^{K}
\frac{\delta Y}{\delta {\cal J}_{\mu } (\omega)
}\frac{d {\cal J}_{\mu}}{d \omega }
 \right]\cr
&=& - i \delta \omega 
\int \frac{d\omega}{2 \pi i}
\left(N\Sigma (\omega)\frac{d g}{d \omega }+K\Pi (\omega)
\frac{d {\cal J}}{d \omega }
 \right)=0
\end{eqnarray}
Integrating this result by parts, we obtain
\begin{equation}\label{valued}
0=\int \frac{d\omega}{2 \pi i}
\left(N g(\omega)\frac{d\Sigma }{d\omega }+K {\cal J} (\omega)
\frac{d \Pi}{d \omega }
 \right)
\end{equation}
This valuable result is a consequence of fermion, or charge
conservation.   

Now the total change in the charge  of the system is given by
\begin{eqnarray}\label{}
\Delta Q &=& N T \sum_{i\omega_{n}}\left[
g (i\omega_{n})
-g_{0} (i
\omega_{n} )\right] e^{i\omega_{n}0^{+}}=
\int \frac{d\omega}{2 \pi i } \left[g (\omega)- g_{0} (\omega)
\right]e^{\omega 0^{+}}
\end{eqnarray}
where the final integral is along the imaginary axis. 

We now subtract the result (\ref{valued}) from this expression, to
obtain
\begin{eqnarray}\label{}
\Delta Q &=& N 
\int \frac{d\omega}{2 \pi i }
\left[
g (\omega)\left(1 -
\frac{d\Sigma }{d\omega} \right) - g_{0} (\omega)\right]e^{\omega 0^{+}} -
 K \int \frac{d\omega}{2 \pi i }
{\cal  J} (\omega)\frac{d \Pi}{d \omega }e^{\omega 0^{+}}\cr
&=&N \int \frac{d\omega}{2 \pi i } 
\frac{d}{d\omega}
\left(
 \ln
[\Sigma - {g}_{0}^{-1} ] -{\rm Tr} \ln
[ - g_{0}^{-1} ] 
 \right)e^{\omega 0^{+}}+
 K \int \frac{d\omega}{2 \pi i }
\frac{d }{d \omega }\ln [-{\cal J}^{-1} (\omega)]e^{\omega 0^{+}}\cr
&=&N \int \frac{d\omega}{2 \pi i } 
\frac{d}{d\omega}
\left(
 \ln
[1 - {g}_{0}\Sigma (\omega) ] 
 \right)e^{\omega 0^{+}}+
 K \int \frac{d\omega}{2 \pi i }
\frac{d }{d \omega }\ln[ -{\cal J}^{-1} (\omega)]e^{\omega 0^{+}}.
\end{eqnarray}
Finally, we distort the contour around the negative imaginary axis, to obtain
\begin{eqnarray}\label{}
\Delta Q &=& \int_{-\infty }^{0}
 \frac{d\omega}{ \pi  } {\rm  Im }
\frac{d}{d\omega}
\left[\phantom{\int} N\ln [1
 - g_{0} (z)\Sigma (z)] 
+K\ln [1 -
J_{0}\Pi (z)]\right]_{z=\omega-i\delta }
\cr 
&=& N \left( \frac{\delta _{c}}{\pi} \right)
+ K \left(\frac{\delta _{\phi }}{\pi} \right)
\end{eqnarray}
where the phase shifts are defined as 
\begin{eqnarray}\label{}
\delta _{c}  &=& {\rm  Im }\left[ \phantom{\int}\hskip -0.4cm
\ln \left(1 - i \pi \rho \Sigma
(\omega-i\delta )\right)\right]_{\omega=0}\cr
\delta _{\phi } &=& {\rm  Im }\left[ \phantom{\int}\hskip -0.4cm
\ln \left( 
1 - J_{0} \Pi
(\omega-i\delta )\right)\right]_{\omega=0}
\end{eqnarray}
Using the compensation theorem to set $\Delta Q=0$ in the infinite
band width limit, we obtain the sum rule 
\begin{equation}\label{}
0= N \left( \frac{\delta _{c}}{\pi} \right)
+ K \left(\frac{\delta _{\phi }}{\pi} \right)
\end{equation}

\section{Appendix B}\label{}

In this section, we prove the finite field Ward Identities, 
\begin{eqnarray}\label{phaseid2}
K \delta_{\phi }+ N \delta _{\downarrow } &=& 0\\
K \delta _{\uparrow} + (N-K)\delta _{\downarrow } &=& K \pi
\end{eqnarray}
The first of these results is a finite field generalization of the
result of Appendix A. In a field, the ``up''  conduction electrons and
the $\phi $ fermion become hybridized, so that 
$G_{\uparrow}$ and ${\cal J}$ are no
longer independent variables. It is convenient to introduce
\[
\tilde{\cal J} (\omega) = \frac{1}{J_{0}^{-1}- \Pi (\omega)}
\]
in terms of which
\begin{eqnarray}\label{links1}
{\cal  J} (\omega) &=& \frac{1}{{\tilde{\cal J}}^{-1} (\omega)-
\tilde{n}_{b}g_{\downarrow } (\omega)}\\ \label{links2}
{g}_{\uparrow} (\omega) &=& \frac{1}{g_{\downarrow }^{-1} (\omega)-
\tilde{n}_{b}\tilde{{\cal J}} (\omega)}
\end{eqnarray}
When we write the Luttinger Ward functional, we must be careful to
express it as a function of independent propagators.  One possible
choice is $g_{\downarrow}$  and ${\cal  J}$. In this case, we can
divide the continuous fermion line running through $Y$ into 
sections which are either $g_{\downarrow }$ or ${\cal  J}$.  
(In this procedure we have effectively integrated out the conduction
electrons  first, so that the $\phi $ fermion propagator contains a
contribution from its hybridization with the conduction electrons in a
field. )
Sections that
involve the ``up'' electron propagator can be broken up into
$g_{\downarrow }$ and ${\cal J}$ using (\ref{links2}).
If we then  write
$Y=Y[g_{\downarrow},{\cal J}]$ and vary the frequency running along the
continuous fermion line, we obtain the finite field version of
(\ref{valued} ), 
\begin{equation}\label{valued2}
\delta Y[g_{\downarrow},{\cal J}] =-i\delta \omega\int \frac{d\omega}{2 \pi i}
\left(N g_{\downarrow }(\omega)\frac{d\Sigma_{\downarrow } }{d\omega }+K {\cal J} (\omega)
\frac{d (\Pi + \tilde{n}_{b}g_{\downarrow })}{d \omega }
 \right)=0
\end{equation}
Following the same steps that were taken in Appendix A, we  obtain the
first of relations (\ref{phaseid2}), 
\[
K \delta_{\phi }+ N \delta _{\downarrow } = 0.
\]
where in a field, 
\begin{eqnarray}\label{}
\delta _{\phi } (B) &=& {\rm  Im }
\left[ \phantom{\int}\hskip -0.4cm
\ln \bigl(
1- J [\Pi(z )+\tilde{n}_{b}g_{\downarrow } (z)] \bigr )
\right]
_{z=0-i\delta }
.\cr\cr
\delta _{c\downarrow  } (B) &=& 
{\rm  Im }\left[\phantom{\int}\hskip -0.4cm 
\ln \bigl ( 1- g_{0}\Sigma_{\downarrow } (z
)\bigr ) \right]
_{z=0-i\delta },
\end{eqnarray}

Now alternatively, we can take the independent propagators to be 
$g_{\uparrow}$, $g_{\downarrow }$ and $\tilde{\cal J}$. That is to
say, we are effectively first integrating out the $\phi $ fermion, so that its
propagator $\cal J$ does not contain a contribution from hybridization with the
``up'' conduction electrons, whereas the  `` up''  electron lines now
contain a contribution due to hybridization with the $\phi $ fermions.
When we make a variation of the frequency along fermion lines inside
$Y$, we obtain
\[
dY[g_{\uparrow},g_{\downarrow },\tilde{{\cal J}}] = i \delta \omega
\int \frac{d\omega}{2 \pi i}
\left[
K \Sigma _{\uparrow} (\omega)\frac{dg_{\uparrow}}{d\omega}
+
(N-K) \Sigma _{\downarrow} (\omega)\frac{dg_{\downarrow}}{d\omega}
+
K
\Pi (\omega)
\frac{d {\tilde{J}}}{d\omega}
 \right]=0,
\]
where $\Sigma _{\uparrow}= \Sigma _{\downarrow }+
\tilde{n}_{b}{\tilde{\cal J}} $, so that 
\begin{eqnarray}\label{}
0 &=& - \int \frac{d\omega}{2 \pi i}
\left[
K g _{\uparrow} (\omega)\frac{d\Sigma _{\uparrow}}{d\omega}
+
(N-K) g _{\downarrow} (\omega)\frac{d\Sigma _{\downarrow}}{d\omega}
+
K{\tilde{J}
} (\omega) \frac{d\Pi }{d\omega}
 \right]\cr
&=& \int \frac{d\omega}{2 \pi i}
\frac{d}{d\omega}\left[ \phantom{\int}\hskip -0.4cm
K \ln \left(1- g _0 \Sigma _{\uparrow} (\omega)
\right)
+ (N-K)\ln \left(1- g _0 \Sigma _{\downarrow}(\omega) \right)
+K \ln  \left(1- J_{0}\Pi(\omega)\right)
 \right]\cr
&=& K\left(\frac{\delta _{\uparrow}}{\pi} \right)
+ (N-K)\left(\frac{\delta _{\downarrow}}{\pi} \right)
+ K \left(\frac{\tilde{\delta }_{\phi }}{\pi} \right)=0
\end{eqnarray}
where we have defined
\[
\tilde{\delta }_{\phi } = {\rm  Im}\ln \left[1 - J_{0}\Pi (z)
\right]\vert _{z= - i \delta }
\]
Now since $\Pi (-i \delta ) = Re \Pi (0)+ i \delta $, $\tilde{\delta
}_{\phi }$ is $0$ or $-\pi$ depending on the sign of $\tilde{J}
(0)$. We may write
\[
K\delta _{\uparrow}
+ (N-K)\delta _{\downarrow}= K (\pi + s)
\]
where $s = \frac{1}{2} (1+ {\rm sgn} (\tilde{\cal J} (0)))= Im \ln (-
\tilde{\cal J} (z ))\vert _{z=-i\delta }$. In actual fact, $\delta
_{\uparrow}$ will jump by $\pi$ exactly at the point where $\tilde{J}$
changes sign, so by redefining
\[
\delta _{\uparrow} = \left. 
{\rm  Im}\ln  \left[\frac{1 - i \pi \rho \Sigma
_{\uparrow} (z)}{- \tilde{{\cal J}} (z)}  \right]\right| _{z= - i \delta }
\]
we obtain a phase shift that evolves smoothly with field, which
satisfies the Ward identity
\[
K\delta _{\uparrow}
+ (N-K)\delta _{\downarrow}= K \pi .
\]


\vfill \eject

\newpage















\end{document}